\documentclass[aps,prb,twocolumn,showpacs,floatfix,superscriptaddress]{revtex4}

\usepackage{amsmath,amssymb}
\usepackage{graphicx}
\usepackage{times}
\usepackage{framed}

\newcommand{\bfB}{{\mathbf{B}}}

\newcommand{\bfsigma}{{\boldsymbol{\sigma}}}

\newcommand{\varF}{{\mathcal{F}}}
\newcommand{\varK}{{\mathcal{K}}}
\newcommand{\varP}{{\mathcal{P}}}
\newcommand{\varQ}{{\mathcal{Q}}}
\newcommand{\varS}{{\mathcal{S}}}
\newcommand{\varU}{{\mathcal{U}}}
\newcommand{\varV}{{\mathcal{V}}}
\newcommand{\varZ}{{\mathcal{Z}}}

\let\lms\leftarrow
\let\rms\rightarrow
\let\up\uparrow
\let\down\downarrow

\newcommand{\Braket}[1]{\mathinner{\langle{\textstyle#1}\rangle}}
\newcommand{\Pde}[2][]{\frac{\partial^{#1}}{\partial{#2}^{#1}}}

\newcommand{\eqnref}[1]{Eq.~(\ref{#1})}
\newcommand{\eqnsref}[1]{Eqs.~(\ref{#1})}
\newcommand{\figref}[1]{Fig.~\ref{#1}}
\newcommand{\Figref}[1]{Figure~\ref{#1}}
\newcommand{\secref}[1]{Sec.~\ref{#1}}
\newcommand{\Secref}[1]{Section~\ref{#1}}

\begin{document}
\title{Current-Conserving Aharonov-Bohm Interferometry with Arbitrary Spin Interactions}
\author{Minchul Lee}
\affiliation{Department of Applied Physics, College of Applied Science, Kyung Hee University, Yongin 446-701, Korea}
\author{Dimitrije Stepanenko}
\affiliation{Department of Physics, University of Basel, Klingelbergstrasse 82,
CH-4056 Basel, Switzerland}

\begin{abstract}
  We propose a general scattering matrix formalism that guarantees the charge
  conservation at junctions between conducting arms with arbitrary spin
  interactions. By using our formalism, we find that the spin-flip scattering
  can happen even at nonmagnetic junctions if the spin eigenstates in arms are
  not orthogonal. We apply our formalism to the Aharonov-Bohm interferometer
  consisting of $n$-type semiconductor ring with both the Rashba spin-orbit
  coupling and the Zeeman splitting. We discuss the characteristics of the
  interferometer as conditional/unconditional spin switch in the
  weak/strong-coupling limit, respectively.
\end{abstract}


\maketitle

\section{Introduction}

Coherent electronic transport through mesoscopic rings or structures with
non-trivial geometries has been extensively investigated both theoretically
\cite{Buttiker1984oct,Loss1990sep,Loss1992jun,Yi1997apr,Romer2000sep,KangK2000dec,Frustaglia2001nov,Meijer2002jul,Hentschel2004apr,Frustaglia2004jun,Wang2005oct,LeeMC2006feb,Lucignano2007jul,Kovalev2007sep,Pletyukhov2008may,Borunda2008dec,Stepanenko2009jun}
and experimentally
\cite{Umbach1984oct,Webb1985jun,Bergsten2006nov,Habib2007apr,Grbic2007oct,Qu2011jun}
in last decades. The studies have aimed at exploring theoretically the quantum
interference in solid-state circuits and also revolutionizing electronic
devices in such a way to exploit the quantum effects. At the heart of studies
of mesoscopic rings, there are two hallmarks of quantum coherence: the
Aharonov-Bohm \cite{Aharonov1959aug} (AB) and Aharonov-Casher
\cite{Aharonov1984jul} (AC) effects. Two effects are related to geometric
phases due to the coupling of a charge to a magnetic flux and of a spin degree
of freedom to an electric field via spin-orbit coupling (SOC),
respectively. Since the AB oscillation in conductance through normal-metal
rings was revealed, \cite{Buttiker1984oct} it is found that the effects can
lead to diverse quantum interference effects such as conductance fluctuations,
\cite{Umbach1984oct} persistent charge and spin current,
\cite{Loss1990sep,Loss1992jun} AB effect for exciton, \cite{Romer2000sep}
mesoscopic Kondo effect, \cite{KangK2000dec} spin switch,
\cite{Frustaglia2001nov,Frustaglia2004jun} spin filter, \cite{LeeMC2006feb} and
spin Hall effect. \cite{Borunda2008dec} From a practical point of view, the
quantum coherent phenomena in mesoscopic rings, especially using the spin
degrees of freedom, have been applied to the fast growing field of spintronics
\cite{Wolf2001nov,Zutic2004apr} and are now known to provide the
easy-to-control devices that generate, manipulate, and detect the
spin-dependent current or signal.

Mesoscopic rings fabricated in semiconductors offer intriguing possibility to
study simultaneously the AB and AC effects because of spin-orbit coupling
naturally formed in crystals. The spin-orbit coupling itself can have various
forms in different materials, leading to diverse current
oscillations.\cite{Frustaglia2004jun,Habib2007apr,Grbic2007oct} In addition,
the strength of the spin-orbit coupling can be controlled by tuning a backgate
voltage to the device. \cite{Nitta1997feb} Among spin-orbit couplings, the
Rashba SOC, originating from the broken structural inversion symmetry, is
linear in momentum and easy to analyze. The studies of spin interference
\cite{Meijer2002jul,Frustaglia2004jun} subject to the Rashba SOC have shown
that the Rashba coupling strength can modulate the unpolarized current,
suggesting the possibility of all-electrical spintronic devices. Recently, a
number of experimental \cite{Habib2007apr,Grbic2007oct} and theoretical
\cite{Kovalev2007sep,Borunda2008dec,Stepanenko2009jun} studies have
investigated transport of heavy holes in rings, whose SOC is cubic in
momentum. In the presence of external magnetic fields, the Zeeman splitting is
operative together with the SOCs and its effect should be taken into
account. \cite{Yi1997apr,Frustaglia2001nov,Hentschel2004apr,Wang2005oct,Lucignano2007jul}



The general framework for the theoretical studies of the mesoscopic transport
relies on the Landauer approach, \cite{Datta1995} and it is described as
tunneling through conduction modes between the source and drain electrodes
coupled to rings.  In the coherent regime, the tunneling is accompanied by
interference between conduction modes. The conductance is then described by a
scattering matrix between modes that carry distinct phases. Due to
interference, the fact that each of the modes is charge conserving does not
guarantee the charge conservation in the total transport. More specifically, a
problem arises when the ring modes form nonorthogonal spin textures; the spins
of the states with the same energy are not orthogonal at every point.
Such a complexity does not arise if the ring has either of the
linear-in-momentum SOCs (like the Rashba SOC) or the Zeeman splitting because
one can then diagonalize the system Hamiltonian such that no mode mixing takes
place. However, in realistic situations with arbitrary form of SOC and Zeeman
splitting, the mode mixing, or spin mixing, naturally exists.


The previous studies considering both the effect of the Rashba SOC and the
Zeeman splitting have dealt with this situation by using the transfer-matrix
method accompanied with wave function matching,
\cite{Yi1997apr,Frustaglia2001nov,Hentschel2004apr} perturbative approach,
\cite{Wang2005oct} and path-integral approach. \cite{Lucignano2007jul} The
transfer-matrix method, even though different group velocities of ring modes
are taken into account, fails guaranteeing the charge conservation at the
lead-ring junction. In their studies, the relation between the lead and the
ring modes are determined solely by the wave function matching, which alone
cannot satisfy the charge conservation. The perturbative calculation is valid
only in the small Zeeman splitting limit. Finally, the path-integral approach,
which may be conceptually useful to interpret the result in terms of phases, is
limited by its semiclassical treatment.

Our goal in this work is to find a general scattering matrix formalism that
guarantees the charge conservation at the lead-ring junctions by its own way of
construction in the presence of arbitrary spin interactions. We detour the mode
mixing problem by introducing artificial spin-independent \textit{buffer}
regions in the vicinity of every junction as shown in \figref{fig:1}. The
mode-mixing effect is then taken care of at the interfaces between the buffers
and the spinful regions in a standard way. Finally, the original system is
recovered in the limit of vanishing buffers.
By using our formalism, we first recover the known results in the case of
orthogonal spin textures and interpret the role of buffers added by
hand. Secondly, we apply our formalism to the $n$-type semiconductor ring with
both the Rashba SOC and the Zeeman splitting. We found that (1) our formalism
truly guarantees the charge conservation at every junction, giving rise to
correct predictions, (2) the spin-flip scattering can happen even if the
junctions are nonmagnetic, (3) the ring interferometer can act as
conditional/unconditional spin switch in the weak/strong-coupling regimes,
respectively, if some conditions are met.

Our paper is structured as follows: Our formalism is introduced and derived in
detail in \secref{sec:formalism}. In \secref{sec:orthogonal} the case of
orthogonal spin texture is treated within our
formalism. \Secref{sec:nonorthogonal} is devoted to the study of a case of
nonorthogonal spin states in which both the Rashba SOC and the Zeeman splitting
are taken into account. Finally, we conclude and summarize our paper in
\secref{sec:discussion}.

\section{General Formalism to Build Current-Conserving Scattering Matrix\label{sec:formalism}}

\subsection{Buffered Structure}

\begin{figure}[!b]
  \centering
  \includegraphics[width=8cm]{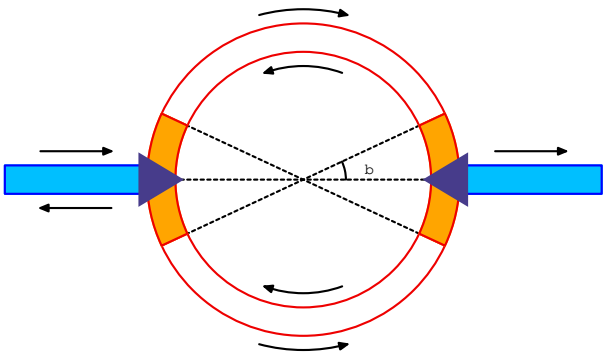}
  \caption{Schematic diagram of buffered Aharonov-Bohm interferometer. The
    upper and low arms of the ring are connected to the junctions through
    buffers whose angular size is given by $\phi_b$.}
  \label{fig:1}
\end{figure}

The scattering in the mesoscopic system is frequently characterized in terms of
the scattering matrix which defines the relative amplitudes and phases of the
scatted states to the injected states. The scattering matrix depends on the
details of the system but is constrained by the laws of conservation. The most
important properties that the scattering matrix obeys are the current
conservation, if there is neither a source or a sink in scatterer and,
additionally the spin current conservation, if the scatterer is
nonmagnetic. The conservation conditions, together with some symmetric
arguments, quite simplify the form of the scattering matrix so that it can be
described by a few parameters. For example, the most frequently used scattering
matrix for the AB interferometer is controlled by a single parameter $\epsilon$
that is varied between 0 (no tunneling) and 1/2 (perfect tunneling). Here the
scattering to and from upper and lower arms is assumed to be symmetric.

This simple construction of the scattering matrix can be extended further to
the magnetic case where spin-dependent interactions exist in the arms. In the
presence of the Zeeman splitting, one can treat the scattering of spin up and
down separately, each of which is described by the simple scattering matrix
mentioned before. For the case of linear-in-momentum spin-orbit coupling such
as the Rashba SOC, one can identify a common spin polarization axis at a
junction so that spin-separate treatment is still possible.  No difficulty in
defining the scattering matrix that satisfies the conservation laws arises as
long as all the arms meeting at a junction share a single spin polarization
axis.  However, the latter condition is quite fragile and generally fails in
the presence of general spin-dependent interactions in the arms. The simplest
case in which it happens is the ring with both the Rashba SOC and the Zeeman
splitting. As will be shown later, the ring eigenstates are neither parallel or
orthogonal to each other in the spin space at all, denying a common spin
polarization axis. The scattering into the ring then involves all the spin
states, preventing spin-separate treatment. The scattering matrix is then
complicated although it should still fulfill the conservation laws.

One can then issue several questions. Is there a general framework to build the
spin-dependent scattering matrix, which guarantees satisfying the conservation
laws by the way of its construction? What is the smallest number of controlling
parameters that are required to describe the scattering in the presence of
arbitrary spin-dependent interactions? Can the spin-flip scattering happen even
if the scatterer itself is still nonmagnetic?

In order to answer these questions we propose a general formalism to build up a
consistent (spin-dependent) scattering matrix for arbitrary spin
interaction. The key idea of our method is to insert artificial buffer regions
between the scatter and arms as depicted in \figref{fig:1}. The buffer regions
are assumed to be free of any spin-dependent interaction. Hence, the scattering
between buffer regions can be described by the simple spin-separate scattering
matrix. The complexity due to the spin-dependent interaction makes its effect
at interfaces between buffers and arms. The wave functions at interfaces are
matched in the systematic way by using the continuity of wave function and its
current density. This wave matching, together with the scattering matrix
between buffer states, leads one to find the scattering matrix between states
of arms. The size of artificial buffers is then shrunken to zero in order to
recover the original configuration. The shrinking does not remove all the
effects of buffers because the effect of scattering at buffer-arm interfaces
still remains. In the end, the scattering matrix connecting states of arms is
constructed.

The advantages of our methods are listed as follows: (1) The scattering matrix
obtained guarantees satisfying the conservation laws. It is because the
scattering between buffer states and the scattering at buffer-arm interfaces
are set up to conserve the charge and spin currents. (2) It systematically
identifies the minimal set of controlling parameters that the scatterer can
have. (3) It provides the reasonable explanation for the effect of spin
interactions in arms on the spin-dependent (possibly spin-flip) scattering even
when the scatterer itself is nonmagnetic.

In the following sections we build up our formalism, especially focusing on the
AB interferometer shown in \figref{fig:1}. The system consists of two leads and
one ring. Each part is assumed to be narrow enough to be regarded as
one-dimensional conductor with a single transverse mode. The scattering matrix
between lead and ring then becomes a $6\times6$ matrix. We will construct a
general scattering matrix for ring with arbitrary spin interaction.  Our
formalism, however, is quite general and can be applied to mesoscopic circuits
with any kind of geometry.

\subsection{Arms: Lead Part}

The leads are composed of normal conductors directing along the $x$
direction. They are free of any magnetic interaction, and their Hamiltonians
read
\begin{align}
  H_{\rm LEAD}
  = \frac{p_x^2}{2m_0^*} + U_0,
\end{align}
where $m_0^*$ is the effective mass of electrons in the leads and $U_0$ is the
minimum energy of the transverse mode. Thanks to the spin degeneracy, the spin
polarization axis of eigenstates can be chosen arbitrarily. Eigenstates of the
leads with the eigenenergy $E$ then are given by
\begin{align}
  e^{\pm iqx} \chi_{\ell\mu}
\end{align}
with the wave number $q = \sqrt{2m_0(E - U_0)}/\hbar$ and the spinors
\begin{align}
  \chi_{\ell+}
  =
  \begin{bmatrix}
    e^{-i\varphi_\ell/2} \cos\vartheta_\ell
    \\
    e^{+i\varphi_\ell/2} \sin\vartheta_\ell
  \end{bmatrix},
  \quad
  \chi_{\ell-}
  =
  \begin{bmatrix}
    - e^{-i\varphi_\ell/2} \sin\vartheta_\ell
    \\
    e^{+i\varphi_\ell/2} \cos\vartheta_\ell
  \end{bmatrix}.
\end{align}
Here $\mu = \pm$ is the spin index, and $\ell = \rm L, R$ the lead index. The
angles $(\vartheta_\ell,\varphi_\ell)$ define the spin polarization axis for
the injection from the left lead ($\ell = \rm L$) and the spin detection axis
in the right lead ($\ell = \rm R$), respectively. In terms of coefficients of
injected ($s_\mu$), reflected ($r_\mu$), and transmitted ($t_\mu$) waves, the
general wave functions in the leads are given by
\begin{subequations}
  \begin{align}
    \nonumber
    \psi_{\rm L}(x)
    & = \sum_\mu [s_\mu e^{iqx} + r_\mu e^{-iqx}] \chi_{\rm L\mu}
    \\
    & = \varU_{\rm L} (e^{iqx} s  + e^{-iqx} r)
    \\
    \psi_{\rm R}(x)
    & = \sum_\mu t_\mu e^{iqx} \chi_{\rm R\mu}
    = \varU_{\rm R} e^{iqx} t
  \end{align}
\end{subequations}
where
\begin{align}
  s
  \equiv
  \begin{bmatrix}
    s_+ \\ s_-
  \end{bmatrix},
  \quad
  r
  \equiv
  \begin{bmatrix}
    r_+ \\ r_-
  \end{bmatrix},
  \quad
  t
  \equiv
  \begin{bmatrix}
    t_+ \\ t_-
  \end{bmatrix}
\end{align}
and
\begin{align}
  \varU_\ell
  \equiv
  \begin{bmatrix}
    e^{-i\varphi_\ell/2} \cos\vartheta_\ell
    & - e^{-i\varphi_\ell/2} \sin\vartheta_\ell
    \\
    e^{+i\varphi_\ell/2} \sin\vartheta_\ell
    & e^{+i\varphi_\ell/2} \cos\vartheta_\ell
  \end{bmatrix}.
\end{align}
The group velocities of the eigenstates are $\pm v_0 \equiv \pm\hbar q/m_0$,
and the charge current densities in the leads are
\begin{align}
  J_{\rm L}
  = v_0 \sum_\mu (|s_\mu|^2 - |r_\mu|^2),
  \quad
  J_{\rm R}
  = v_0 \sum_\mu |t_\mu|^2.
\end{align}

\subsection{Arms: Ring Part}

Ring can be either of normal conductor, $n$-type semiconductor, or $p$-type
semiconductor. It is narrow enough that the radial dimension is constant with
the radius $\rho_0$ and the degree of freedom is solely described by the
azimuthal angle $\phi$. We assume that an external magnetic field $\bfB$ is
applied so that the ring encloses a magnetic flux $\Phi$, or the dimensionless
flux $f = \Phi/\Phi_0$ with the flux quantum $\Phi_0 = hc/e$ and the spin
splitting arises due to the Zeeman term
\begin{align}
  H_Z = \frac{g^*\mu_B}{2} \bfsigma\cdot\bfB,
\end{align}
where $g^*$ is the Land\'e $g$-factor, $\mu_B$ the Bohr magneton, and
$\bfsigma$ the Pauli matrices. For semiconductor rings, an appropriate
spin-orbit interaction $H_{\rm SO}$ is operative. The ring Hamiltonian is then
given by
\begin{align}
  \label{eq:Hring}
  H_{\rm RING}
  =
  E_0 (-i\partial_\phi - f)^2 + H_{\rm SO} + \frac{g^*\mu_B}{2} \bfsigma\cdot\bfB
\end{align}
with
\begin{align}
  E_0 = \frac{\hbar^2}{2m^*\rho_0^2}
\end{align}
where $m^*$ is the effective mass in the ring. In general the Hamiltonian has
four eigenstates labeled by the spin index $\mu = \pm$ and the propagation
direction $\varrho = +$ (counterclockwise) and $-$ (clockwise) for each energy
$E$. Each eigenstate is endowed with a wave number $k_\mu^\varrho$, the
solution of the dispersion relation. The wave number can be real (propagating
state) or complex number (evanescent wave). The general form of the eigenstates
is then written as
\begin{align}
  \varphi_\mu^\varrho(\phi)
  =
  e^{i(k_\mu^\varrho + f)\phi}
  \begin{bmatrix}
    a_\mu^\varrho(\phi)
    \\
    b_\mu^\varrho(\phi)
  \end{bmatrix}.
\end{align}
In terms of coefficients $u_\mu^\varrho$ for the upper arm (U) and
$d_\mu^\varrho$ for the lower arm (D), the ring wave functions are given by
\begin{subequations}
  \label{eq:wftn:ring}
  \begin{align}
    \psi_{\rm U}(\phi)
    & = \sum_{\mu\varrho} u_\mu^\varrho \varphi_\mu^\varrho(\phi)
    = \sum_\varrho \varU^\varrho(\phi) \varK^\varrho(\phi) u^\varrho
    \\
    \psi_{\rm D}(\phi)
    & = \sum_{\mu\varrho} d_\mu^\varrho \varphi_\mu^\varrho(\phi)
    = \sum_\varrho \varU^\varrho(\phi) \varK^\varrho(\phi) d^\varrho,
  \end{align}
\end{subequations}
where
\begin{align}
  u^\varrho
  \equiv
  \begin{bmatrix}
    u_+^\varrho \\ u_-^\varrho
  \end{bmatrix},
  \quad
  d^\varrho
  \equiv
  \begin{bmatrix}
    d_+^\varrho \\ d_-^\varrho
  \end{bmatrix}
\end{align}
and
\begin{align}
  \varU^\varrho(\phi)
  \equiv
  \begin{bmatrix}
    a_+^\varrho(\phi) & a_-^\varrho(\phi)
    \\
    b_+^\varrho(\phi) & b_-^\varrho(\phi)
  \end{bmatrix},
  \quad
  \varK^\varrho(\phi)
  \equiv
  e^{if\phi}
  \begin{bmatrix}
    e^{ik_+^\varrho\phi} & 0
    \\
    0 & e^{ik_-^\varrho\phi}
  \end{bmatrix}.
\end{align}
The group velocity of each eigenstate, $\varrho v_\mu^\varrho$ is given by the
expectation value $\Braket{\varphi_\mu^\varrho|v_\phi|\varphi_\mu^\varrho}$ of
the velocity operator $v_\phi$. Note that the spin-orbit interaction affects
the velocity operator, and in general the energy eigenstate is not the
eigenstate of the velocity operator. If the time reversal symmetry is not
broken, the relations $v_+^+ = v_-^-$ and $v_+^- = v_-^+$ hold generally no
matter what the spin-orbit interaction is. In terms of the group velocities,
the charge current densities in the ring are expressed as
\begin{subequations}
  \begin{align}
    J_{\rm U}
    & = \sum_\mu (v_\mu^+ |u_\mu^+|^2 - v_\mu^- |u_\mu^-|^2)
    \\
    J_{\rm D}
    & = \sum_\mu (v_\mu^+ |d_\mu^+|^2 - v_\mu^- |d_\mu^-|^2)
  \end{align}
\end{subequations}
for the upper and lower arms, respectively. For later use, we define the wave
function applied by the velocity operator
\begin{subequations}
  \label{eq:vwftn:ring}
  \begin{align}
    v_\phi\psi_{\rm U}(\phi)
    & = \sum_\varrho \varrho \varV^\varrho(\phi) \varK^\varrho(\phi) u^\varrho
    \\
    v_\phi\psi_{\rm D}(\phi)
    & = \sum_\varrho \varrho \varV^\varrho(\phi) \varK^\varrho(\phi) d^\varrho,
  \end{align}
\end{subequations}
where the $2\times2$ matrix $\varV^\varrho(\phi)$ depends on the details of the
system.

\subsection{Buffers}

In our formalism, no buffer region is inserted between the leads and the
junctions. It is because the leads are free of spin-dependent interaction like
buffers and the scattering at the interface between the lead and the buffer
becomes trivial. On the other hand, as shown in \figref{fig:1}, the buffer
regions are inserted between the junctions and the ring arms. In the AB
interferometer, therefore, four buffer regions with the same angular size
$\phi_b$ are defined in the left/right side of upper/lower arms. Having the
junctions at the angles $\phi_{\rm L}$ and $\phi_{\rm R} = 0$, the interfaces
between the buffers and the arms are located at $\phi_{\rm UR} = \phi_b$,
$\phi_{\rm UL} = \phi_{\rm L} - \phi_b$, $\phi_{\rm DL} = \phi_{\rm L} +
\phi_b$, and $\phi_{\rm DR} = 2\pi - \phi_b$. Since the buffers are free of any
spin-dependent interaction like leads, the Hamiltonian in buffers reads
\begin{align}
  H_{\rm BUFFER}
  = E_0 (-i\partial_\phi - f)^2 + U_b,
\end{align}
where $U_b$ is the offset in the band bottom with respect to the ring
part. With no spin interaction, it is free to choose the spin polarization axes
in them, and the axis of each buffer is chosen to that of the
nearest-neighboring lead. Eigenstates of the buffers with the energy $E$,
labeled by $\mu$ and $\varrho$, are then given by
\begin{align}
  e^{i(\varrho\kappa + f)\phi} \chi_{\ell\mu}
\end{align}
with the wave number $\kappa = \sqrt{(E - U_b)/E_0}$ and the nearby-lead index
$\ell$. By defining the coefficients $u_{\ell\mu}^\varrho$ for the upper buffers
close to the side $\ell$ and $d_{\ell\mu}^\varrho$ for the lower buffers
close to the side $\ell$, the buffer wave functions are written as
\begin{subequations}
  \label{eq:wftn:buffer}
  \begin{align}
    \psi_{\rm U\ell}(\phi)
    & =
    \sum_{\mu\varrho} u_{\ell\mu}^\varrho e^{i(\varrho\kappa+f)(\phi-\phi_\ell)}
    \chi_{\ell\mu}
    = \sum_\varrho \varU_\ell \varK_b^\varrho(\phi) u_\ell^\varrho
    \\
    \psi_{\rm D\ell}(\phi)
    & =
    \sum_{\mu\varrho} d_{\ell\mu}^\varrho e^{i(\varrho\kappa+f)(\phi-\phi_\ell)}
    \chi_{\ell\mu}
    = \sum_\varrho \varU_\ell \varK_b^\varrho(\phi) d_\ell^\varrho
  \end{align}
\end{subequations}
where
\begin{align}
  u_\ell^\varrho
  \equiv
  \begin{bmatrix}
    u_{\ell+}^\varrho \\ u_{\ell-}^\varrho
  \end{bmatrix},
  \qquad
  d_\ell^\varrho
  \equiv
  \begin{bmatrix}
    d_{\ell+}^\varrho \\ d_{\ell-}^\varrho
  \end{bmatrix}
\end{align}
and
\begin{align}
  \varK_b^\varrho
  \equiv
  e^{i(\varrho\kappa + f)(\phi-\phi_\ell)}
  \begin{bmatrix}
    1 & 0
    \\
    0 & 1
  \end{bmatrix}.
\end{align}
The group velocities of the eigenstates are simply given by $\pm v_b \equiv
\pm\hbar \kappa/m\rho_0$, which is the eigenvalue of the velocity operator
$v_\phi = (\hbar/m\rho_0)(-i\partial_\phi - f)$, and the charge current
densities in the buffers are
\begin{subequations}
  \begin{align}
    J_{\rm U\ell}
    & = v_b \sum_\mu (|u_{\ell\mu}^+|^2 - |u_{\ell\mu}^-|^2)
    \\
    J_{\rm D\ell}
    & = v_b \sum_\mu (|d_{\ell\mu}^+|^2 - |d_{\ell\mu}^-|^2)
  \end{align}
\end{subequations}
for the left/right and upper/lower buffers, respectively. For later use, we
define the wave function applied by the velocity operator
\begin{subequations}
  \label{eq:vwftn:buffer}
  \begin{align}
    v_\phi\psi_{\rm U\ell}(\phi)
    & = v_b \varU_\ell (\varK_b^+ u_\ell^+ - \varK_b^- u_\ell^-)
    \\
    v_\phi\psi_{\rm D\ell}(\phi)
    & = v_b \varU_\ell (\varK_b^+ d_\ell^+ - \varK_b^- d_\ell^-).
  \end{align}
\end{subequations}

\subsection{Lead-Buffer Scattering Matrices}

With the buffered structure, the scattering at the junctions connects the
states in the leads and the buffers. Since both the leads and the buffers have
no magnetic interaction, the conventional scattering matrix can be defined to
describe the scattering at the junctions. The reasonable conditions for the
scattering matrix are that (1) no spin flip takes place, (2) the scatterings
from and to the upper and lower arms are same, (3) no phase shift is acquired,
and (4) the charge current is conserved. The first condition makes the
scattering matrix diagonal in the spin space, and due to the second condition
the scattering matrix with respect to the normalized flux is symmetric in the
exchange between the upper and lower arms. The most general lead-buffer
scattering matrix satisfying the above conditions is then
\begin{align}
  \varS
  =
  \begin{bmatrix}
    \varS_{11} & \varS_{12}
    \\
    \varS_{21} & \varS_{22}
  \end{bmatrix}
  =
  \begin{bmatrix}
    \varS_{0,11} \otimes \sigma_0 & \varS_{0,12} \otimes \sigma_0
    \\
    \varS_{0,21} \otimes \sigma_0 & \varS_{0,22} \otimes \sigma_0
  \end{bmatrix}
\end{align}
with
\begin{subequations}
  \label{eq:S0}
  \begin{align}
    \varS_{0,11}
    & = -\zeta \sqrt{1 - 2\epsilon}
    \\
    \varS_{0,12}
    & =
    \sqrt{\frac{v_b}{v_0}\epsilon}
    \begin{bmatrix}
      1 & 1
    \end{bmatrix}
    \\
    \varS_{0,12}
    & =
    \sqrt{\frac{v_0}{v_b}\epsilon}
    \begin{bmatrix}
      1 \\ 1
    \end{bmatrix}
    \\
    \varS_{0,22}
    & =
    \begin{bmatrix}
      \frac{\zeta}{2} (\sqrt{1 - 2\epsilon} - 1)
      & \frac{\zeta}{2} (1 + \sqrt{1 - 2\epsilon})
      \\
      \frac{\zeta}{2} (1 + \sqrt{1 - 2\epsilon})
      & \frac{\zeta}{2} (\sqrt{1 - 2\epsilon} - 1)
    \end{bmatrix}
  \end{align}
\end{subequations}
with $\zeta = \pm$. Here the controlling parameter $\epsilon$ varies from 0
(perfect transmission) to 1/2 (complete decoupling), and $\sigma_0$ is
$2\times2$ identity matrix, indicating the absence of spin-flip
scattering. Throughout this paper, we set $\zeta = +1$ considering the case of
phase-conserving scattering between upper and lower arms.

Assuming that both the junctions have the same scattering matrix, one can set
up linear equations for the coefficients of lead and buffer states: at the
left junction
\begin{subequations}
  \label{eq:sm:buffer:left}
  \begin{align}
    r & = \varS_{11} s + \varS_{12} c_{\rm L}^\lms
    \\
    c_{\rm L}^\rms & = \varS_{12} s + \varS_{22} c_{\rm L}^\lms
  \end{align}
\end{subequations}
and at the right junction
\begin{subequations}
  \label{eq:sm:buffer:right}
  \begin{align}
    t & = \varS_{12} c_{\rm R}^\rms
    \\
    c_{\rm R}^\lms & = \varS_{22} c_{\rm R}^\rms,
  \end{align}
\end{subequations}
where the left- and right-moving buffer states are defined as
\begin{align}
  c_\ell^\lms
  \equiv
  \begin{bmatrix}
    u_\ell^+ \\ d_\ell^-
  \end{bmatrix}
  =
  \begin{bmatrix}
    u_{\ell+}^+ \\ u_{\ell-}^+ \\ d_{\ell+}^- \\ d_{\ell-}^-
  \end{bmatrix},
  \quad
  c_\ell^\rms
  \equiv
  \begin{bmatrix}
    u_\ell^- \\ d_\ell^+
  \end{bmatrix}
  =
  \begin{bmatrix}
    u_{\ell+}^- \\ u_{\ell-}^- \\ d_{\ell+}^+ \\ d_{\ell-}^+
  \end{bmatrix},
\end{align}
respectively.

Note that the form of the scattering matrix guarantees the charge and spin
current conservation by the way of its construction.

\subsection{Lead-Arm Scattering Matrices}

Now we derive the scattering matrix connecting the lead states and the ring
states. To do that, we need to find out the linear relations between the buffer
states and the ring states. The relations are to be determined from the
boundary conditions at the interfaces by using the continuity of wave function
$\psi(\phi)$ and the current conservation. The latter condition can be
reformulated in terms of the continuity of $H(\phi) \psi(\phi)$ where $H(\phi)$
is the Hamiltonian defined simultaneously in the buffer and ring regions. As
long as the spin-orbit interaction is composed of the linear and/or second
orders of the momentum operator, the continuity of $H(\phi) \psi(\phi)$ leads
to the continuity of $v_\phi \psi(\phi)$. Now we apply the boundary conditions
at four interfaces. By using \eqnsref{eq:wftn:ring} and (\ref{eq:wftn:buffer}),
the continuity of the wave function, $\psi_{\rm U}(\phi_{\rm U\ell}) =
\psi_{\rm U\ell}(\phi_{\rm U\ell})$ and $\psi_{\rm D}(\phi_{\rm D\ell}) =
\psi_{\rm D\ell}(\phi_{\rm D\ell})$ at the interfaces gives rise to
\begin{subequations}
  \begin{align}
    \sum_\varrho
    \varU^\varrho(\phi_{\rm U\ell}) \varK^\varrho(\phi_{\rm U\ell}) u^\varrho
    & =
    \varU_\ell
    \sum_\varrho \varK_b^\varrho(\phi_{\rm U\ell}) u_\ell^\varrho
    \\
    \sum_\varrho
    \varU^\varrho(\phi_{\rm D\ell}) \varK^\varrho(\phi_{\rm D\ell}) d^\varrho
    & =
    \varU_\ell
    \sum_\varrho \varK_b^\varrho(\phi_{\rm D\ell}) d_\ell^\varrho.
  \end{align}
\end{subequations}
The second continuity conditions, $v_\phi \psi_{\rm U}(\phi_{\rm U\ell}) =
v_\phi \psi_{\rm U\ell}(\phi_{\rm U\ell})$ and $v_\phi \psi_{\rm D}(\phi_{\rm
  D\ell}) = v_\phi \psi_{\rm D\ell}(\phi_{\rm D\ell})$, together with
\eqnsref{eq:vwftn:ring} and (\ref{eq:vwftn:buffer}), lead to
\begin{subequations}
  \begin{align}
    \sum_\varrho
    \varrho \varV^\varrho(\phi_{\rm U\ell}) \varK^\varrho(\phi_{\rm U\ell}) u^\varrho
    & =
    v_b \varU_\ell
    \sum_\varrho \varrho \varK_b^\varrho(\phi_{\rm U\ell}) u_\ell^\varrho
    \\
    \sum_\varrho
    \varrho \varV^\varrho(\phi_{\rm D\ell}) \varK^\varrho(\phi_{\rm D\ell}) d^\varrho
    & =
    v_b \varU_\ell
    \sum_\varrho \varrho \varK_b^\varrho(\phi_{\rm D\ell}) d_\ell^\varrho.
  \end{align}
\end{subequations}
It is straightforward to solve the equations for the coefficients of the buffer
states:
\begin{subequations}
  \label{eq:bufferringrelation}
  \begin{align}
    u_\ell^\varrho
    & =
    [\varK_b^\varrho(\phi_{\rm U\ell})]^{-1}
    \sum_{\varrho'}
    \varZ_\ell^{\varrho\varrho'}(\phi_{\rm U\ell})
    \varK^{\varrho'}(\phi_{\rm U\ell})
    u^{\varrho'}
    \\
    d_\ell^\varrho
    & =
    [\varK_b^\varrho(\phi_{\rm D\ell})]^{-1}
    \sum_{\varrho'}
    \varZ_\ell^{\varrho\varrho'}(\phi_{\rm D\ell})
    \varK^{\varrho'}(\phi_{\rm D\ell})
    d^{\varrho'}
  \end{align}
\end{subequations}
with
\begin{align}
  \label{eq:Z}
  \varZ_\ell^{\varrho\varrho'}(\phi)
  \equiv
  \varU_\ell^{-1}
  \frac{\varU^{\varrho'}(\phi) + \varrho \varV^{\varrho'}(\phi)/v_b}{2}.
\end{align}
Once the relations between coefficients of the buffer and the ring states are
set up, it is time to shrink the buffers by setting $\phi_b\to0$. The buffer
propagating matrices, $\varK_b^\varrho$ become the identity matrix just because
of zero propagating distance. Here some caution should be made about the limit
values of the interface points. The left interfaces go to the single point,
$\phi_{\rm UL}, \phi_{\rm DL} \to \phi_{\rm L} \equiv \phi_{\rm L}^\pm$, while
the limit values of the right interfaces are different, $\phi_{\rm UR} \to 0
\equiv \phi_{\rm R}^+$ and $\phi_{\rm DR} \to 2\pi \equiv \phi_{\rm R}^-$.

Combining \eqnsref{eq:sm:buffer:left}, (\ref{eq:sm:buffer:right}), and
(\ref{eq:bufferringrelation}), one can build linear equations for the
coefficients of lead and ring states, which are similar to
\eqnsref{eq:sm:buffer:left} and (\ref{eq:sm:buffer:right}): at the left
junction
\begin{subequations}
  \label{eq:sm:ring:left}
  \begin{align}
    r & = \varS_{\rm L,11} s + \varS_{\rm L,12} \varK_{\rm L}^\lms c^\lms
    \\
    \varK_{\rm L}^\rms c^\rms
    & = \varS_{\rm L,12} s + \varS_{\rm L,22} \varK_{\rm L}^\lms c^\lms
  \end{align}
\end{subequations}
and at the right junction
\begin{subequations}
  \label{eq:sm:ring:right}
  \begin{align}
    t & = \varS_{\rm R,12} \varK_{\rm R}^\rms c^\rms
    \\
    \varK_{\rm R}^\lms c^\lms & = \varS_{\rm R,22} \varK_{\rm R}^\rms c^\rms,
  \end{align}
\end{subequations}
where the left- and right-moving ring states and the propagating matrices are
defined as
\begin{align}
  c^\lms
  \equiv
  \begin{bmatrix}
    u^+ \\ d^-
  \end{bmatrix}
  =
  \begin{bmatrix}
    u_+^+ \\ u_-^+ \\ d_+^- \\ d_-^-
  \end{bmatrix},
  \quad
  c^\rms
  \equiv
  \begin{bmatrix}
    u^- \\ d^+
  \end{bmatrix}
  =
  \begin{bmatrix}
    u_+^- \\ u_-^- \\ d_+^+ \\ d_-^+
  \end{bmatrix},
\end{align}
and
\begin{align}
  \varK_\ell^\lms
  \equiv
  \begin{bmatrix}
    \varK^+(\phi_\ell^+)) &
    \\
    & \varK^-(\phi_\ell^-)
  \end{bmatrix},
  \quad
  \varK_\ell^\rms
  \equiv
  \begin{bmatrix}
    \varK^-(\phi_\ell^+)) &
    \\
    & \varK^+(\phi_\ell^-)
  \end{bmatrix},
\end{align}
respectively. The lead-ring scattering matrices
\begin{align}
  \varS_\ell
  =
  \begin{bmatrix}
    \varS_{\ell,11} & \varS_{\ell,12}
    \\
    \varS_{\ell,21} & \varS_{\ell,22}
  \end{bmatrix}
\end{align}
are then given by
\begin{subequations}
  \label{eq:S}
  \begin{align}
    \varS_{\ell,11}
    & =
    \varS_{11}
    + \varS_{12}
    \varQ_\ell^- (\varQ_\ell^+ - \varS_{22} \varQ_\ell^-)^{-1} \varS_{21}
    \\
    \varS_{\ell,12}
    & =
    \varS_{12}
    \left[
      \varP_\ell^+
      + \varQ_\ell^- (\varQ_\ell^+ - \varS_{22} \varQ_\ell^-)^{-1}
      (\varS_{22} \varP_\ell^+ - \varP_\ell^-)
    \right]
    \\
    \varS_{\ell,21}
    & = (\varQ_\ell^+ - \varS_{22} \varQ_\ell^-)^{-1} \varS_{21}
    \\
    \varS_{\ell,22}
    & =
    (\varQ_\ell^+ - \varS_{22} \varQ_\ell^-)^{-1}
    (\varS_{22} \varP_\ell^+ - \varP_\ell^-)
  \end{align}
\end{subequations}
with
\begin{subequations}
  \label{eq:PQ}
  \begin{align}
    \varP_{\rm L}^\varrho
    & \equiv
    \begin{bmatrix}
      \varZ_{\rm L}^{\varrho+}(\phi_{\rm L}) &
      \\
      & \varZ_{\rm L}^{\varrho-}(\phi_{\rm L})
    \end{bmatrix},
    &
    \varQ_{\rm L}^\varrho
    & \equiv
    \begin{bmatrix}
      \varZ_{\rm L}^{\varrho-}(\phi_{\rm L}) &
      \\
      & \varZ_{\rm L}^{\varrho+}(\phi_{\rm L})
    \end{bmatrix}
    \\
    \varP_{\rm R}^\varrho
    & \equiv
    \begin{bmatrix}
      \varZ_{\rm R}^{\varrho-}(\phi_{\rm R}^+) &
      \\
      & \varZ_{\rm R}^{\varrho+}(\phi_{\rm R}^-)
    \end{bmatrix},
    &
    \varQ_{\rm R}^\varrho
    & \equiv
    \begin{bmatrix}
      \varZ_{\rm R}^{\varrho+}(\phi_{\rm R}^+) &
      \\
      & \varZ_{\rm R}^{\varrho-}(\phi_{\rm R}^-)
    \end{bmatrix}.
  \end{align}
\end{subequations}
From \eqnsref{eq:S} and (\ref{eq:PQ}), a few immediate general features of the
scattering matrix can be discussed: (1) In general, the matrices
$\varZ_\ell^{\varrho\varrho'}$ are not spin diagonal. It means that the
lead-ring scattering matrices are not diagonal in the spin basis even if we
start with the assumption that the junction itself does not invoke the
spin-flip scattering. For example, the spin up injected from the lead can be
reflected into the spin down for any spin injection axis. It is not because the
junction is a magnetic scatterer but because of spin-dependent interaction in
the ring. The magnetic property in the arms of the ring can invoke the
spin-dependent scattering at the junctions. (2) The buffer effect remains. The
lead-ring scattering matrix have two controlling parameters: $\epsilon$ and
$U_b$. The latter parameter enters into the scattering matrix in terms of the
buffer group velocity $v_b$. The velocity $v_b$ appears in the scattering
matrix in two ways: in the overall factor $\sqrt{v_b/v_0}$ of $\varS_{12}$ and
$\varS_{21}$ [see \eqnref{eq:S0}] and in the matrices
$\varZ_\ell^{\varrho\varrho'}$ [see \eqnref{eq:Z}]. The overall factor
$\sqrt{v_b/v_0}$ appears in $\varS_{\ell,ij}$ in the same way as in
$\varS_{ij}$ and does not affect the spin-dependent scattering discussed
above. On the other hand, $v_b$ in the matrices $\varZ_\ell^{\varrho\varrho'}$
can tune magnitudes of its off-diagonal components. Therefore, we can draw a
conclusion that at least two parameters for junctions, here $\epsilon$ and
$U_b$, are necessary to specify and control the spin-dependent scattering due
to arbitrary spin-dependent interaction in arms.

We'd like to emphasize that the scattering matrix, \eqnref{eq:S} is the only
solution that guarantees the conservation of the charge and spin currents at
junctions under our symmetric assumptions. Since we have used the simplest
buffer structure that introduces only one additional parameter, more
complexity, if necessary, can be introduced into the scattering matrix by
allowing additional interactions in the buffer. Here we introduce the minimal
scattering matrix working properly in the presence of general spin-orbit
interaction.

\subsection{Reflection and Transmission Coefficients}

It is now quite straightforward to solve \eqnsref{eq:sm:ring:left} and
(\ref{eq:sm:ring:right}) in order to obtain the spin-resolved reflection and
transmission coefficients in terms of the lead-ring scattering matrix
$\varS_{\ell,ij}$:
\begin{subequations}
  \begin{align}
    t
    & =
    \varS_{\rm R,12}
    \left(
      \varK^\rms \varF
      -
      \varS_{\rm L,22} \varK^\lms \varF \varS_{\rm R,22}
    \right)^{-1}
    \varS_{\rm L,21} s
    \\
    \nonumber
    r
    & =
    \left[
      \varS_{\rm L,11}
      +
      \varS_{\rm L,12} \varK^\lms \varF \varS_{\rm R,22}
    \right.
    \\
    & \qquad\quad\left.\mbox{}
      \times
      \left(
        \varK^\rms \varF
        -
        \varS_{\rm L,22} \varK^\lms \varF \varS_{\rm R,22}
      \right)^{-1}
      \varS_{\rm L,21}
    \right] s
  \end{align}
\end{subequations}
with
\begin{subequations}
  \begin{align}
    \varK^\lms
    & =
    {\rm diag}
    \left(
      e^{ik_+^+\phi_{\rm L}},
      e^{ik_-^+\phi_{\rm L}},
      e^{-ik_+^-(2\pi-\phi_{\rm L})},
      e^{-ik_-^-(2\pi-\phi_{\rm L})}
    \right)
    \\
    \varK^\rms
    & \equiv
    {\rm diag}
    \left(
      e^{ik_+^-\phi_{\rm L}},
      e^{ik_-^-\phi_{\rm L}},
      e^{-ik_+^+(2\pi-\phi_{\rm L})},
      e^{-ik_-^+(2\pi-\phi_{\rm L})}
    \right)
    \\
    \varF
    & \equiv
    {\rm diag}
    \left(
      e^{if\phi_{\rm L}},
      e^{if\phi_{\rm L}},
      e^{-if(2\pi-\phi_{\rm L})},
      e^{-if(2\pi-\phi_{\rm L})}
    \right).
  \end{align}
\end{subequations}
Note that the overall factors $\sqrt{v_b/v_0}$ in $\varS_{\ell,12}$ and
$\varS_{\ell,21}$ are canceled out in the reflection and transmission
coefficients. Therefore, the velocity in the leads does not affect the
coefficients at all.

Below we calculate the transmission amplitudes $T_{\mu\mu'} = |t_{\mu\mu'}|^2$,
and by using them the charge conductance
\begin{align}
  G = \frac{e^2}{h} \sum_{\mu\mu'} T_{\mu\mu'}
\end{align}
and the current polarization
\begin{align}
  P = \frac12 \sum_{\mu\mu'} \mu T_{\mu\mu'}.
\end{align}
with respect to unpolarized input current are obtained.

\section{Orthogonal Spin States\label{sec:orthogonal}}

Before proceeding to study the case in which our formalism is indispensable, we
want to apply it to the simple cases where the spin-separate treatment is
possible. As mentioned in the previous section, the spin-separate treatment can
be used when the ring is of the normal conductor, or has the linear-in-momentum
spin-orbit coupling such as Rashba SOC, or has the Zeeman splitting only. What
is in common in all the cases is that the group velocity and the spin matrix
are direction-independent, $v_\mu^\varrho = v_\mu$ and $\varU^\varrho(\phi) =
\varU(\phi)$ and that the energy eigenstates are also the eigenstates of the
corresponding velocity operator,
\begin{align}
  v_\phi \varphi_\mu^\varrho(\phi)
  = \varrho v_\mu \varphi_\mu^\varrho(\phi)
\end{align}
($v_+ \ne v_-$ only when the Zeeman splitting exists).  Then the matrix
$\varV^\rho(\phi)$ in \eqnref{eq:vwftn:ring} is simply given by
\begin{align}
  \varV^\rho(\phi)
  =
  \varU(\phi)
  \begin{bmatrix}
    v_+ & 0
    \\
    0 & v_-
  \end{bmatrix}.
\end{align}
Accordingly, the matrices $\varZ_\ell^{\varrho\varrho'}(\phi)$ are simplified to
\begin{align}
  \label{eq:Z:oss}
  \varZ_\ell^{\varrho\varrho'}(\phi)
  =
  [\varU_\ell^{-1} \varU(\phi)]
  \begin{bmatrix}
    z_+^\varrho & 0
    \\
    0 & z_-^\varrho
  \end{bmatrix}
\end{align}
with
\begin{align}
  z_\mu^\varrho
  \equiv \frac{1 + \varrho v_\mu/v_b}{2}.
\end{align}
By setting $\varU_\ell = \varU(\phi_\ell)$ (note that $\varU(\phi_{\rm R}^+)$
and $\varU(\phi_{\rm R}^-)$ usually differ only up to the overall phase
factor), the matrices $\varZ_\ell^{\varrho\varrho'}(\phi_\ell)$ become spin
diagonal, and consequently we recover the spin-separate lead-ring scattering
matrix. For each spin component, the lead-ring scattering matrix for spin $\mu$
can be expressed as
\begin{subequations}
  \begin{align}
    \varS_{\ell\mu,11}
    & =
    \varS_{\mu,11}
    + \varS_{\mu,12} z_\mu^- (z_\mu^+ - z_\mu^- \varS_{\mu,22})^{-1} \varS_{\mu,21}
    \\
    \varS_{\ell\mu,12}
    & =
    \varS_{\mu,12}
    \left[
      z_\mu^+
      + z_\mu^-
      (z_\mu^+ - z_\mu^- \varS_{\mu,22})^{-1} (z_\mu^+ \varS_{\mu,22} - z_\mu^-)
    \right]
    \\
    \varS_{\ell\mu,21}
    & = (z_\mu^+ - z_\mu^- \varS_{\mu,22})^{-1} \varS_{\mu,21}
    \\
    \varS_{\ell\mu,22}
    & = (z_\mu^+ - z_\mu^- \varS_{\mu,22})^{-1} (z_\mu^+ \varS_{\mu,22} - z_\mu^-).
  \end{align}
\end{subequations}
The buffer effect due to the velocity mismatch at buffer-ring interfaces still
remains in the above expressions. However, one can recover the original form of
the scattering matrix by redefining the controlling parameter $\epsilon$. In
other words, one can easily prove that the above scattering matrix can be
rewritten as
\begin{align}
  \varS_{\ell\mu,ij}(\epsilon,U_b) = \varS_{ij}(\epsilon'_\mu)
\end{align}
with
\begin{align}
  \label{eq:e}
  \epsilon'_\mu(\epsilon,U_b)
  =
  \frac{(v_\mu/v_b) \epsilon}{(z_\mu^+ + \zeta z_\mu^- \sqrt{1 - 2\epsilon})^2}.
\end{align}
Note that $0 \le \epsilon'_\mu \le 1/2$ for $0 \le \epsilon \le 1/2$ and $0 <
v_b < \infty$, as expected. It implies that in the cases where the
spin-separate treatment is possible the only role of the buffer is to
renormalize the tunneling parameter $\epsilon$ through \eqnref{eq:e}. Hence the
buffer is unnecessary and the junction can be characterized by a single
parameter $\epsilon'_\mu$ of arbitrary values. However, our formalism reveals
the possible origin of spin-dependent values for $\epsilon'_\mu$. The
difference between $\epsilon'_\mu$ for two spins is due to different group
velocity $v_\mu$ in the ring and consequent difference in the magnitude of
velocity mismatch at the junction. Even though it is convention in literature
to define a same value of $\epsilon$ for two spins, it is more physically
correct to have different tunneling parameters for two spin components, as
shown in our formalism.

\section{Nonorthogonal Spin States\label{sec:nonorthogonal}}

As an application of our formalism, we consider the $n$-type semiconductor ring
with both the Rashba SOC and the Zeeman splitting. First, we set up the
lead-arm scattering matrix in this case and then examine the features of the
scattering matrix. After that, the spin-resolved transport through the ring is
investigated.

\subsection{Setup of Scattering Matrix}

The Rashba spin-orbit interaction in the ring geometry is given by
\begin{align}
  \begin{split}
    H_{\rm SO}
    & =
    \frac{\alpha}{\rho_0}
    \left[
      (\sigma_x \cos\phi + \sigma_y \sin\phi) \left(-i\Pde{\phi} - f\right)
    \right.
    \\
    & \qquad\qquad\left.\mbox{}
      +
      \frac{i}{2} (\sigma_x \sin\phi - \sigma_y \cos\phi)
    \right].
  \end{split}
\end{align}
It is straightforward to calculate the eigenstates of the ring Hamiltonian,
\eqnref{eq:Hring} and we obtain, for a given energy $E \ge
E_+(\gamma_R,\gamma_Z)$, four eigenstates\cite{eigenstate_validity}
\begin{subequations}
  \label{eq:evec}
  \begin{align}
    \psi_+^\varrho(\phi)
    & =
    e^{i(k_+^\varrho + f)\phi}
    \begin{bmatrix}
      e^{-i\phi/2} \cos\frac{\theta_+^\varrho}{2}
      \\
      e^{+i\phi/2} \sin\frac{\theta_+^\varrho}{2}
    \end{bmatrix}
    \\
    \psi_-^\varrho(\phi)
    & =
    e^{i(k_-^\varrho + f)\phi}
    \begin{bmatrix}
      - e^{-i\phi/2} \sin\frac{\theta_-^\varrho}{2}
      \\
      e^{+i\phi/2} \cos\frac{\theta_-^\varrho}{2}
    \end{bmatrix},
  \end{align}
\end{subequations}
where the wave numbers are the solutions of
\begin{align}
  \label{eq:k}
  \frac{E}{E_0}
  =
  [k_\mu^\varrho]^2
  + \mu \sqrt{(\gamma_Z - k_\mu^\varrho)^2 + (\gamma_R k_\mu^\varrho)^2}
  + \frac14
\end{align}
with dimensionless constants
\begin{align}
  \gamma_Z
  \equiv \frac{g^*\mu_BB/2}{E_0}
  \quad\text{and}\quad
  \gamma_R
  \equiv \frac{\alpha/\rho_0}{E_0}.
\end{align}
Here $E_+(\gamma_R,\gamma_Z)$ is the energy bottom of the upper spin branch
($\mu = +$), and the angles are defined via
\begin{subequations}
  \label{eq:tiltangle}
  \begin{align}
    \cos\theta_\mu^\varrho
    & =
    \frac{\gamma_Z - k_\mu^\varrho}%
    {\sqrt{(\gamma_Z - k_\mu^\varrho)^2 + (\gamma_R k_\mu^\varrho)^2}}
    \\
    \sin\theta_\mu^\varrho
    & =
    \frac{\gamma_R k_\mu^\varrho}%
    {\sqrt{(\gamma_Z - k_\mu^\varrho)^2 + (\gamma_R k_\mu^\varrho)^2}}.
  \end{align}
\end{subequations}
Note that the spin textures of the eigenstates are all crownlike as in the
Rashba SOC-only case: The effective magnetic field for each eigenstate has the
radial and $z$-directional components whose relative strength is determined by
the angle $\theta_\mu^\varrho$. However, in this case, the angles
$\theta_\mu^\varrho$ are all different, which may lead to complicated
(energy-dependent) spin precession along the ring. On the reversal of the Zeeman
splitting, \eqnsref{eq:k} and (\ref{eq:tiltangle}) guarantees the following
relations:
\begin{align}
  \label{eq:ksym}
  k_\mu^\varrho(\gamma_Z) = -k_\mu^{\bar\varrho}(-\gamma_Z)
  \quad\text{and}\quad
  \theta_\mu^\varrho(\gamma_Z) = \theta_\mu^{\bar\varrho}(-\gamma_Z) + \mu\pi.
\end{align}

Both the parameters $\gamma_Z$ and $f$ are proportional to the magnetic field
$B$, and their ratio is fixed to
\begin{align}
  \label{eq:ratio}
  \frac{\gamma_Z}{f} = g^* \frac{m^*}{m},
\end{align}
where $m$ is the electron mass in vacuum.  In solids, the effective mass of
electrons can be much smaller than its raw value. So the dimensionless flux $f$
can vary over successive integers with a negligible change in $\gamma_Z$.

In order to build the lead-ring scattering matrices, \eqnref{eq:S}, one needs
to construct the appropriate matrices $\varU^\rho(\phi)$ and
$\varV^\varrho(\phi)$. By using the above eigenstates and the velocity operator
\begin{align}
  v_\phi
  =
  \frac{\hbar}{m\rho_0} \left(- i\Pde{\phi} - f\right)
  +
  \frac{\alpha}{\hbar} (\sigma_x \cos\phi + \sigma_y \sin\phi),
\end{align}
the matrices for the $n$-type semiconductor ring are found to be
\begin{widetext}
  \begin{subequations}
    \begin{align}
      \varU^\rho(\phi)
      & =
      \begin{bmatrix}
        e^{-i\phi/2} \cos\frac{\theta_+^\varrho}{2}
        & -e^{-i\phi/2} \sin\frac{\theta_-^\varrho}{2}
        \\
        e^{+i\phi/2} \sin\frac{\theta_+^\varrho}{2}
        & e^{+i\phi/2} \cos\frac{\theta_-^\varrho}{2}
      \end{bmatrix}
      \\
      \varV^\rho(\phi)
      & =
      \varrho \frac{\hbar}{m\rho_0}
      \left(
        \begin{bmatrix}
          k_+^\varrho e^{-i\phi/2} \cos\frac{\theta_+^\varrho}{2}
          & - k_-^\varrho e^{-i\phi/2} \sin\frac{\theta_-^\varrho}{2}
          \\
          k_+^\varrho e^{+i\phi/2} \sin\frac{\theta_+^\varrho}{2}
          & k_-^\varrho e^{+i\phi/2} \cos\frac{\theta_-^\varrho}{2}
        \end{bmatrix}
        +
        \frac{1}{2\cos\theta_R}
        \begin{bmatrix}
          - e^{-i\phi/2} \cos\frac{2\theta_R-\theta_+^\varrho}{2}
          & - e^{-i\phi/2} \sin\frac{2\theta_R-\theta_-^\varrho}{2}
          \\
          - e^{+i\phi/2} \sin\frac{2\theta_R-\theta_+^\varrho}{2}
          & e^{+i\phi/2} \cos\frac{2\theta_R-\theta_-^\varrho}{2}
        \end{bmatrix}
      \right)
    \end{align}
  \end{subequations}
\end{widetext}
with the Rashba angle $\theta_R$ defined via
\begin{align}
  \cos\theta_R
  \equiv
  - \frac{1}{\sqrt{1 + \gamma_R^2}}
  \quad\text{and}\quad
  \sin\theta_R
  \equiv
  \frac{\gamma_R}{\sqrt{1 + \gamma_R^2}}.
\end{align}
These matrices enter into \eqnref{eq:Z} and determine the lead-arm scattering
matrices in \eqnref{eq:S} once the injection and detection spin axes are fixed
through $\varU_\ell$.

For a closed ring, the single-valued condition quantizes the ring levels:
\begin{align}
  \label{eq:ringlevel}
  n = k_\mu^\varrho(E) + f - \frac12,
\end{align}
where $n$ is any integer.

\subsection{Lead-Arm Scattering Matrix}

\begin{figure}[!t]
  \centering
  \includegraphics[width=6cm]{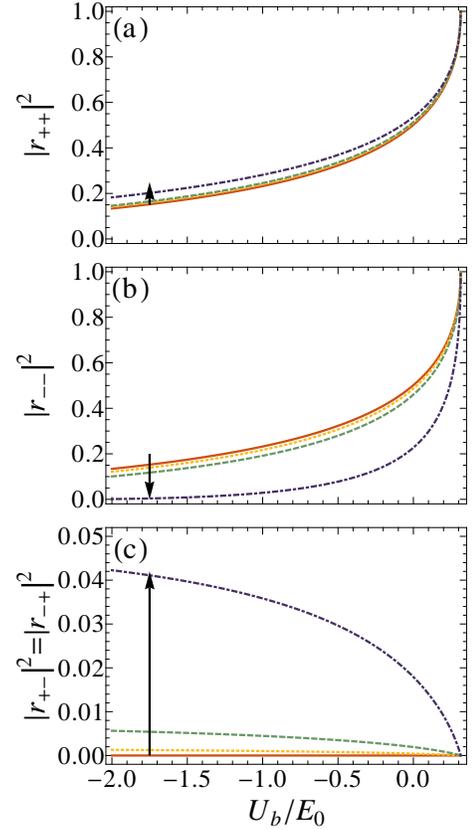}
  \caption{(color online) Reflection amplitudes as functions of $U_b(\le E)$
    for different values of $\gamma_Z$: 0 (solid), $\gamma_R/2$ (dotted),
    $\gamma_R$ (dashed), and $2\gamma_R$ (dot-dashed). Here we set $\epsilon =
    1/4$, $\gamma_R = 0.1$, and $E = 1.02\times
    E_+(\gamma_R=0.1,\gamma_Z=0.2)$. The spin polarization in the lead is set
    to be along the $x$ axis: $(\vartheta_{\rm L},\varphi_{\rm L}) =
    (\pi/2,0)$. The arrows indicate the trend with increasing $\gamma_Z$.}
  \label{fig:2}
\end{figure}

In this section we examine the matrix elements of lead-arm $S$-matrix,
\eqnref{eq:S} in the presence of both the Rashba SOC and Zeeman terms. For
later use, the matrix elements of $S$-matrix for the left junction ($\ell = \rm
L$) are named as
\begin{align}
  \varS_{\rm L,11}
  =
  \begin{bmatrix}
    r_{++} & r_{+-}
    \\
    r_{-+} & r_{--}
  \end{bmatrix}
  \ \text{and}\quad
  \varS_{\rm L,21}
  =
  \begin{bmatrix}
    t_{u++} & t_{u+-}
    \\
    t_{u-+} & t_{u--}
    \\
    t_{d++} & t_{d+-}
    \\
    t_{d-+} & t_{d--}
  \end{bmatrix}.
\end{align}
First, we focus on the spin-flip scattering taking place in the lead
side. \Figref{fig:2} shows the dependence of the reflection amplitudes
$|r_{\mu\mu'}|^2$ on $U_b$ for different values of $\gamma_Z$ with $\gamma_R$
being fixed at a finite value. In the absence of the Zeeman splitting
$(\gamma_Z = 0)$, we obtain $|r_{++}|^2 = |r_{--}|^2$ and $|r_{+-}|^2 =
|r_{-+}|^2 = 0$ as expected. We numerically confirmed that this is true
regardless of the polarization axis $(\vartheta_\ell, \varphi_\ell)$, the
Rashba SOC strength $\gamma_R$, the junction parameters $\epsilon$ and
$U_b$. That is, no spin-flip reflection takes place when only the Rashba SOC
exists. In this case the role of $U_b$ is to simply renormalize $\epsilon$ [see
\eqnref{eq:e}] as displayed in \figref{fig:2}(a) and (b): The perfect
transmission can happen at some values of $U_b$ even though $\epsilon = 1/4 <
1/2$ is used.

\begin{figure}[!t]
  \centering
  \includegraphics[width=7cm]{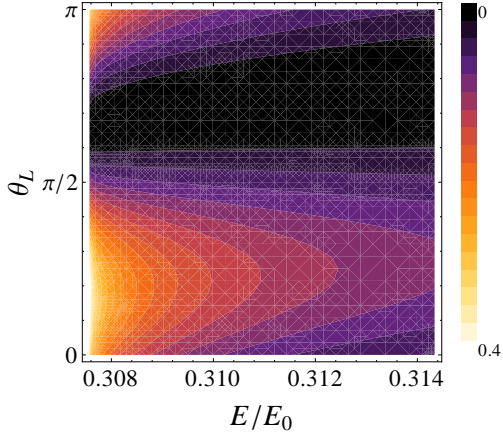}
  \caption{(color online) Contour plot of spin-flip reflection amplitude
    $|r_{+-}|^2$ as a function of $E$ and $\vartheta_{\rm L}$. Here we set
    $\epsilon = 1/2$, $U_b = 0$, $\gamma_R = 0.1$, $\gamma_Z = 2\gamma_R$,
    $\varphi_{\rm L} = 0$. The energy ranges from
    $E_+(\gamma_R=0.1,\gamma_Z=0.2)$ to $1.02 E_+(\gamma_R=0.1,\gamma_Z=0.2)$.}
  \label{fig:3}
\end{figure}

\begin{figure}[!t]
  \centering
  \includegraphics[width=8.5cm]{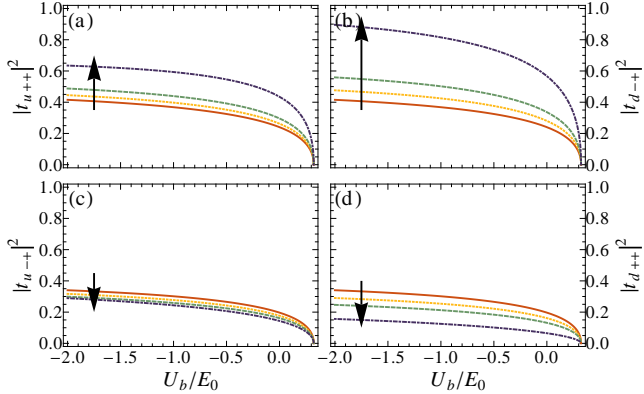}
  \caption{(color online) Transmission amplitudes $|t_{u\mu+}|^2$ and
    $|t_{d\mu+}|^2$ as function of $U_b$ for spin $+$ injection from the
    lead. Values of parameters and plot styles are same as in \figref{fig:2}
    except $E = 1.05\times E_+(\gamma_R=0.1,\gamma_Z=0.2)$.}
  \label{fig:4}
\end{figure}

On the other hand, the spin-conserving feature of the reflection is no longer
valid as soon as the Zeeman splitting is switched on. \Figref{fig:2}(c) clearly
shows that the spin-flip reflection occurs for finite values of $\gamma_Z$ and
its amplitude, $|r_{+-}|^2 = |r_{+-}|^2$ increases with $\gamma_Z$. The
spin-flip reflection depends sensitively on the incident energy $E$ and the
polarization axis $(\vartheta_\ell,\varphi_\ell)$ as wells as $U_b$, as can be
seen in \figref{fig:3}. It modulates with the spin polarization axis in the
lead, and more importantly, decreases rapidly with increasing $E$. Although the
amplitude of spin-flip scattering can be considerable close to the band bottom,
$E_+$, it becomes negligibly small with the incident energy $E$ well above the
band bottom. It explains why the previous works
\citep{Yi1997apr,Frustaglia2001nov,Hentschel2004apr} could not notice the
breakdown of the current conservation with their wrong $S$-matrix: Unless the
energy is close to the band bottom, the spin-flip scattering makes quite small
contribution to the total current. However, its presence, though being small,
is important to fulfill both the current conservation and the correct mathcing
of the wave function.

\Figref{fig:4} displays the transmission amplitudes $|t_{u\mu+}|^2$ and
$|t_{d\mu+}|^2$ as functions of $U_b$ for spin $\mu=+$ injection from the
lead. In the absence of the Zeeman splitting, $|t_{u++}|^2 = |t_{d-+}|^2$ and
$|t_{u-+}|^2 = |t_{d++}|^2$ hold no matter what values the other parameters
have. Similar relations can be found for spin $-$ injection as wells. It is
because the eigenstates $\varphi_+^+(\phi)$ and $\varphi_+^-(\phi)$ make
time-reversal pairs with $\varphi_-^-(\phi)$ and $\varphi_-^+(\phi)$,
respectively. However, the introduction of finite Zeeman splitting breaks the
time reversal symmetry of the system, and the balance between the transmission
coefficients is gone. The transmission amplitudes for different $\mu$ and
$\varrho$ behave differently with increasing $\gamma_Z$ because the group
velocities $v_\mu^\varrho$ are all different and the spin overlap between the
injected wave and the eigenstates also get different from each other. Note that
the transmission amplitudes are not necessarily smaller than one since it is
the current, not the tunneling coefficient that satisfies the unitary condition.

\begin{figure}[!t]
  \centering
  \includegraphics[width=7cm]{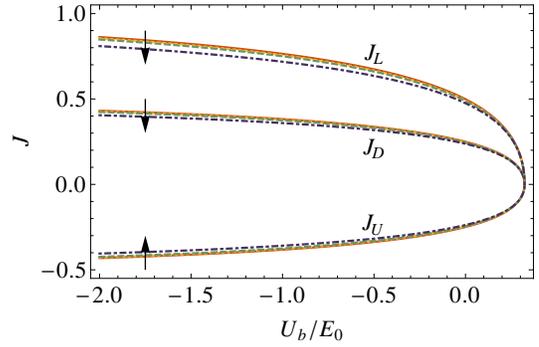}
  \caption{(color online) Charge currents, $J_{\rm L}$, $J_{\rm U}$, and
    $J_{\rm D}$ as function of $U_b$ with respect to a unit spin $+$ polarized
    current $(v_0 = 1)$ from the lead. Values of parameters and plot styles are
    same as in \figref{fig:4}.}
  \label{fig:5}
\end{figure}

\begin{figure}[!t]
  \centering
  \includegraphics[width=5.5cm]{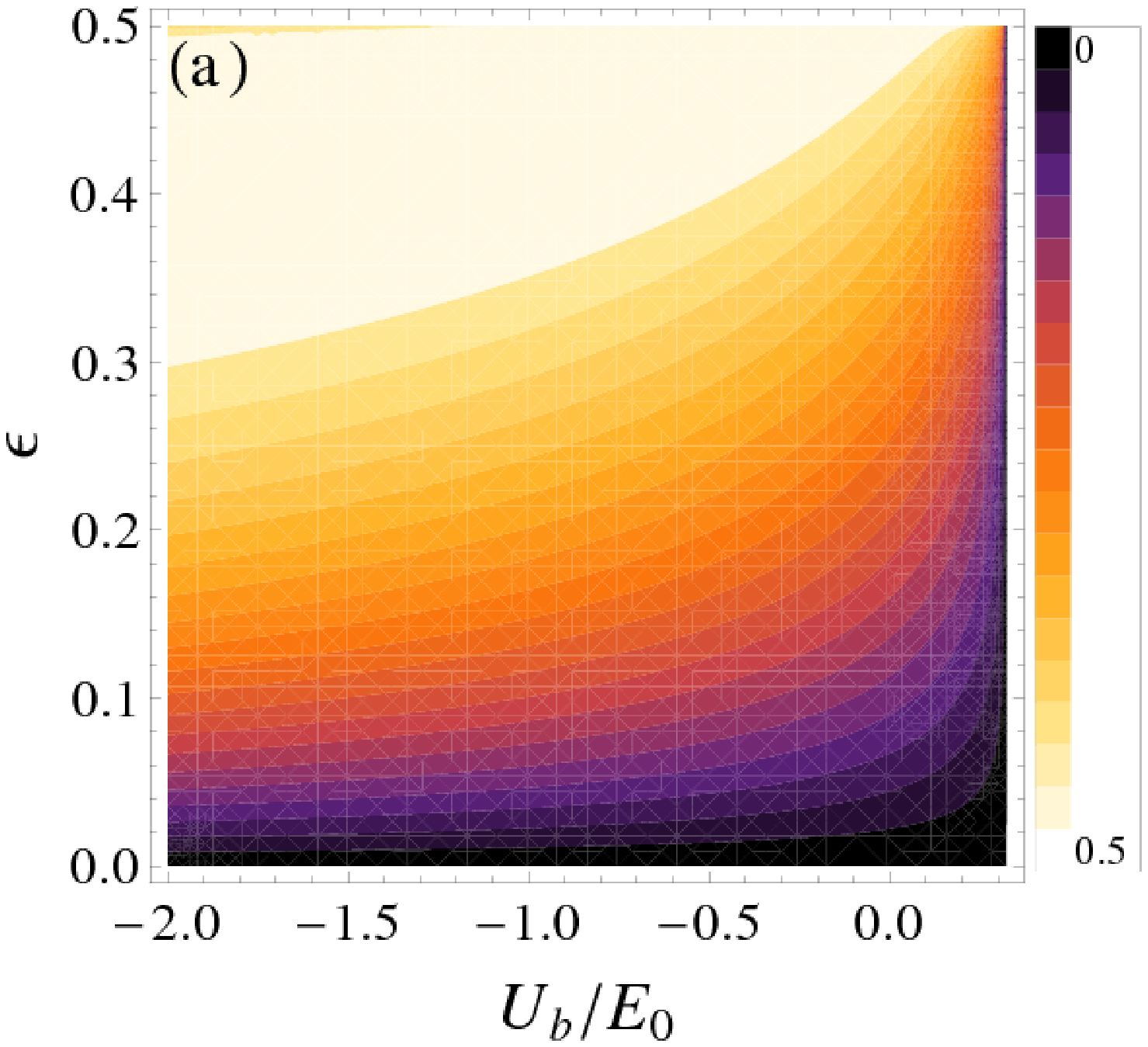}\\
  \includegraphics[width=5.5cm]{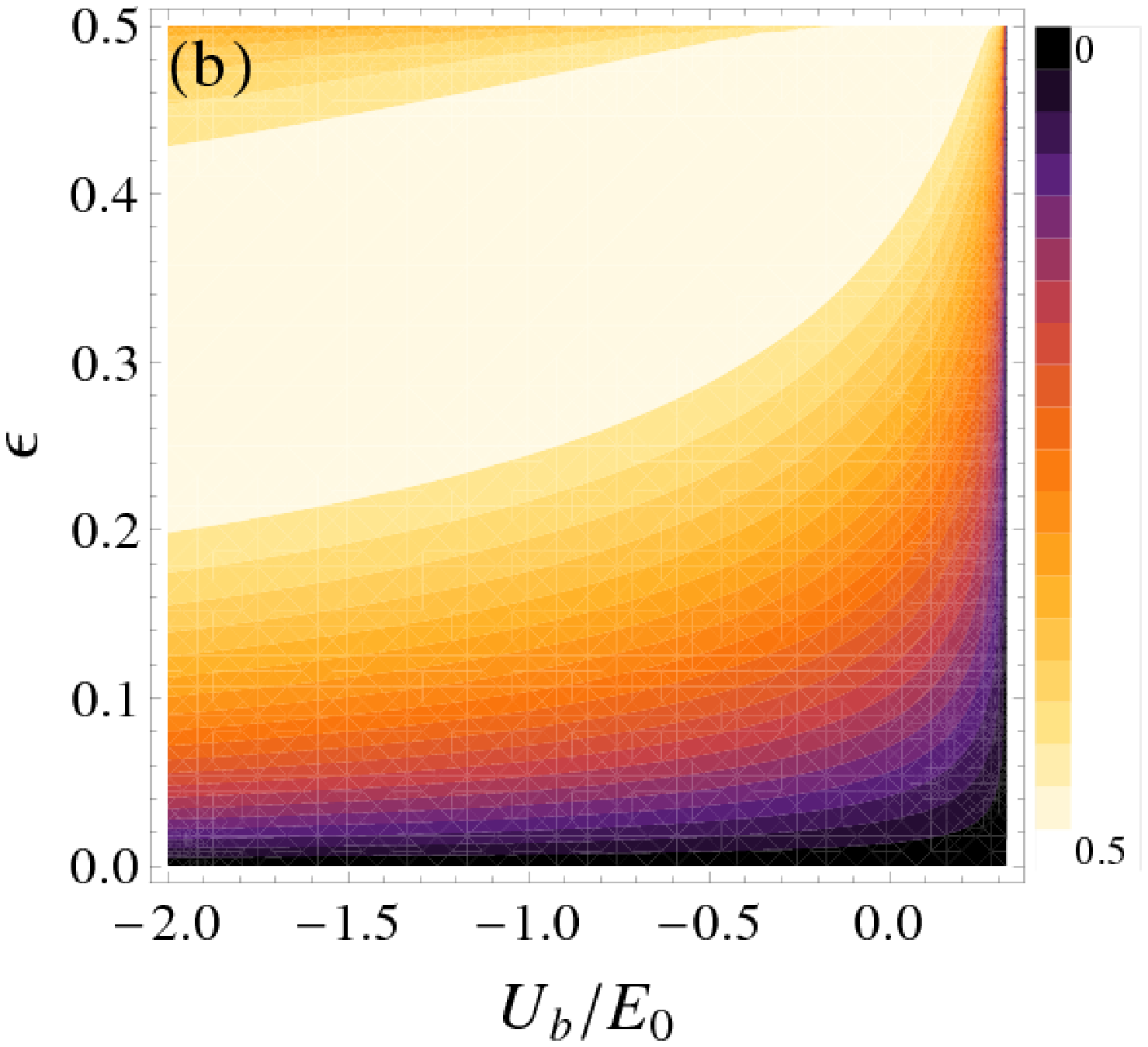}\\
  \caption{(color online) Contour plots of effective control parameters
    $\epsilon_+$ [(a)] and $\epsilon_-$ [(b)] as functions of $\epsilon$ and
    $U_b$ for $\gamma_R = 0.1$, $\gamma_Z = 2\gamma_R$, $E = 1.05\times
    E_+(\gamma_R=0.1,\gamma_Z=0.2)$, and $(\vartheta_{\rm L},\varphi_{\rm L}) =
    (\pi/2,0)$.}
  \label{fig:6}
\end{figure}

As proposed in our formalism, the charge current conservation, $J_{\rm L} +
J_{\rm U} - J_{\rm D} = 0$ is well satisfied as shown in \figref{fig:5}.
Interestingly, the time-reversal breaking and its consequences on the
transmission amplitudes do not invalidate the symmetric scattering to two arms
imposed on the raw $S$-matrix, \eqnref{eq:S0}. As can be seen from
\figref{fig:5}, the normalized currents in both arms are observed to always
satisfy $J_{\rm U} = - J_{\rm D}$. One would guess that the imbalances in the
transmission amplitudes [see \figref{fig:4}] and non-orthogonality of the
eigenstates lead to the asymmetry between the scatterings at upper and lower
buffer-arm interfaces. However, our calculations show that the symmetric
property of the junction remains untouched based on the fact that the raw
$S$-matrix is symmetric and the upper and lower arms are identical.

Finally, we extract the effective spin-dependent control parameters
$\epsilon_\mu$ from
\begin{align}
  \epsilon_\mu
  \equiv
  \frac{1 - \sum_{\mu'} |r_{\mu'\mu}|^2}{2}
\end{align}
as a function of $\epsilon$ and $U_b$ in \figref{fig:6}. As expected,
$\epsilon_\mu$ depends sensitively on $U_b$, and is spin-dependent: $\epsilon_+
\ne \epsilon_-$. Moreover, it also depends on the ring property such as the
strength of Rashba SOC and Zeeman term so that in contrast to the conventional
scattering theory the scattering at a junction is not determined solely by the
junction itself but is affected by the arm property as wells.

\subsection{Aharonov-Bohm Interferometry}

In this section we investigate the charge and spin transport through the
Aharonov-Bohm type interferometer in the presence of both the Rashba SOC and
Zeeman terms. We divide the study into two regimes: weak- and strong-coupling
limits. In the weak-coupling regime where the effective control parameters
$\epsilon_\mu$ are small, the transport features the quantized levels in the
ring, while in the strong-coupling regime the interference between the
eigenstates is important.

\subsubsection{Weak-Coupling Limit}

\begin{figure}[!t]
  \centering
  \includegraphics[width=7.5cm]{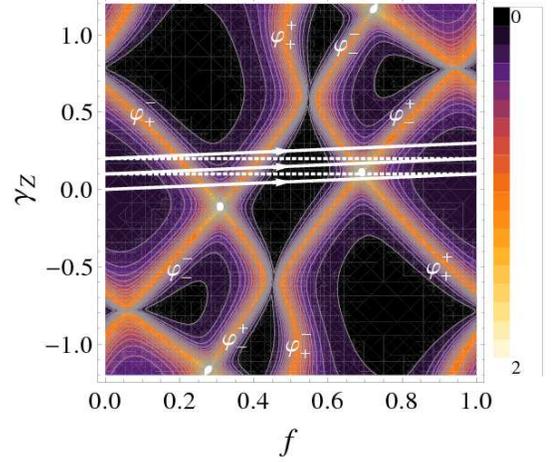}
  \caption{(color online) Contour plot of charge conductance $G$ in unit of
    $e^2/h$ as a function of $f$ and $\gamma_Z$ in the weak coupling limit with
    $\epsilon = 0.15$ and $U_b = 0$. Here we have used $\gamma_R = 0.4$ and $E
    = 2 E_+(\gamma_R=0.4,\gamma_Z=0.8)$. The white lines follow the linear
    relation between $f$ and $\gamma_Z$: $\gamma_Z = 0.1\times f$.}
  \label{fig:7}
\end{figure}

\begin{figure}[!t]
  \centering
  \includegraphics[width=8cm]{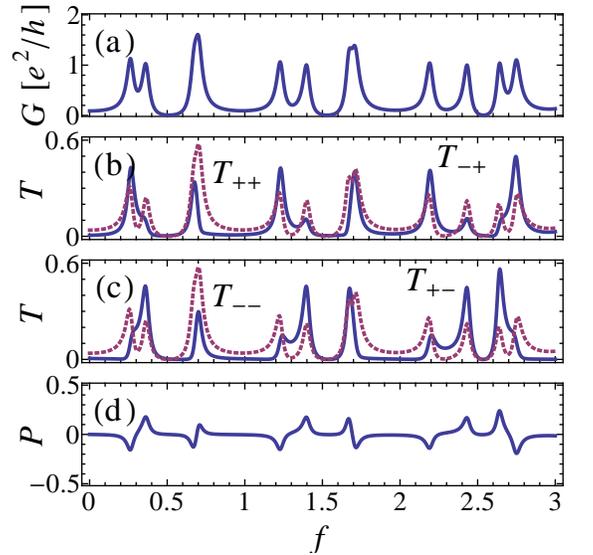}
  \caption{(color online) (a) Charge conductance $G$, (b,c) spin-conserving
    transmission amplitudes $T_{++}$, $T_{--}$ (dotted lines) and spin-flip
    transmission amplitudes $T_{-+}$, $T_{+-}$ (solid lines), and (d) current
    polarization as functions of $f$ along the white lines in \figref{fig:7}
    with $\gamma_Z = 0.1\times f$. Here the polarization axis of two leads are
    chosen to align with the positive $x$ axis: $(\vartheta_\ell,\varphi_\ell)
    = (\pi/2,0)$. Values of other parameters are same as in \figref{fig:7}.}
  \label{fig:8}
\end{figure}

\Figref{fig:7} shows a typical dependence of the charge conductance $G$ on $f$
and $\gamma_Z$ in the weak-coupling limit with $\epsilon = 0.15$ and $U_b =
0$. The high transmission (the bright lines) occurs when the quantization
condition, \eqnref{eq:ringlevel} is satisfied. Here the resonant tunneling via
the quantized ring levels boosts the transmission. This boosting is not
affected by the choice of the spin polarization axis in the leads. Exactly same
charge conductance is obtained by taking the spin polarization axis along the
$z$ axis instead of the $x$ axis used in \figref{fig:7}.
The conductance plot is symmetric with respect to the point
$(f,\gamma_Z)=(0,0)$, which is attributed to the relations in \eqnref{eq:ksym}.
In addition, the resonance lines exhibit the anti-crossing-like behavior, which
is absent in the quantized levels themselves, \eqnref{eq:ringlevel}. The
anti-crossing behavior originates from the Fano-like anti-resonance between two
degenerate ring states whose spin polarizations are rather parallel, leading to
large overlap between their wavefunctions. In this case, the injected state
with any spin polarization has almost same overlaps with the degenerate ring
states, resulting destructive interference between them in the transmitted
state. It happens mostly when the time-reversal pair states
$(\varphi_+^+,\varphi_-^-)$ or $(\varphi_-^+,\varphi_+^-)$ cross as seen in
\figref{fig:7} and less frequently when the counter-propagating pair states
$(\varphi_+^+,\varphi_+^-)$ or $(\varphi_-^+,\varphi_-^-)$ do. For the pairs
$(\varphi_+^+, \varphi_-^+)$ or $(\varphi_+^-, \varphi_-^-)$, their spin
polarizations are almost orthogonal to each other so that the transport through
each state is almost independent of that through the other, and their
transmission amplitudes are simply additive.

In \figref{fig:8}(a) the charge conductance is calculated as a function of
external magnetic field $B$ or the normalized flux $f$ by taking into account
the linear relation, \eqnref{eq:ratio} between $\gamma_Z$ and $f$ with the
ratio $g^* m^*/m = 0.1$ which is indicated by the white lines in
\figref{fig:7}. The charge conductance clearly exhibits four (or three) peaks
as the magnetic flux is increased by one flux quantum $\Phi_0$.  The accidental
degeneracy in the ring levels enhances the conductance further, while it is
still smaller than the two-channel maximum value $2e^2/h$. The fluctuations in
the peak heights is mainly due to the variation of spin polarization axis of
the ring eigenstates at junctions.

Each ring eigenstate, having the crownlike spin texture, brings about the
spin-flip transport as shown in \figref{fig:8}(b) and (c). While the peaks in
the spin-flip transmission amplitudes (solid lines) are located at the same
positions as those in the charge conductance, they alternate between $T_{+-}$
and $T_{-+}$: the $\mu=+$ level give rise to the enhancement of $T_{-+}$ and
the $\mu=-$ level to that of $T_{+-}$. This level dependence is easily
understood from the fact that the spin polarization of the $\mu=+/-$ level has
inward/outward radial component. Since the tilt angle $\theta_\mu^\varrho$
varies between 0 and $\pi$, however, the spin-flip amplitudes also fluctuate.
In addition, each level also makes the comparable contribution to the
spin-conserving transmissions, $T_{++} = T_{--}$ (dotted lines), which follow
the behavior of the charge conductance.
These spin-dependent transmission enables the unpolarized current input to
generate the spin polarized current. As seen in \figref{fig:8}(d), the current
polarization $P$ exhibits peaks and valleys whenever the spin-flip transmission
is enhanced. However, since the spin blocking or the spin flip occur only
partially, its magnitude is usually much smaller than 1/2.

\begin{figure}[!t]
  \centering
  \includegraphics[width=8cm]{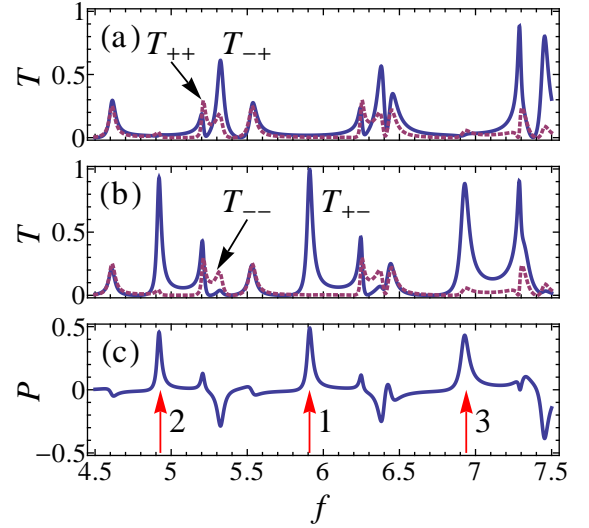}
  \caption{(color online) (a,b) Spin-conserving transmission amplitudes
    $T_{++}$, $T_{--}$ (dotted lines) and spin-flip transmission amplitudes
    $T_{-+}$, $T_{+-}$ (solid lines), and (c) current polarization as functions
    of $f$ with the relation $\gamma_Z = 0.1\times f$. The condition $k^+_+ =
    \gamma_Z$ is exactly satisfied at the point 1, and the energy $E$ is given
    by \eqnref{eq:E2} with respect to the solutions of \eqnref{eq:ringlevel2}
    for $n=6$.  Here we have used $\epsilon = 0.1$, $U_b = 0$ and $\gamma_R =
    0.4$. The red arrows indicates the points (1,2,3) where the spin switch is
    close to its maximum.}
  \label{fig:9}
\end{figure}
In order to achieve the complete spin polarization or spin flip, the lead spin
axis should be set to align with the (energy-dependent) spin polarization of
the level at the junction. However, the arbitrary tuning of the spin
polarization of the lead is not easy to implement. Instead, one can adjust the
spin polarization of the ring level to the predefined spin axis of the lead by
tuning the external magnetic field. The formulas for the tilt angle,
\eqnref{eq:tiltangle} show that a special adjustment, $k_\mu^\varrho =
\gamma_Z$ yields $\theta_\mu^\varrho = \pm\pi/2$, setting the spin polarization
axis of the arm state at junctions along the $x$ direction. The adjustment
requires the energy
\begin{align}
  \label{eq:E2}
  \frac{E}{E_0} = \gamma_Z^2 + |\gamma_Z \gamma_R| + \frac14
\end{align}
(here $\mu = +$ is chosen) and the quantization condition
\begin{align}
  \label{eq:ringlevel2}
  n = \gamma_Z + f - \frac12.
\end{align}
From \eqnref{eq:ringlevel2}, together with \eqnref{eq:ratio}, the candidates
for the magnetic field $B$ or the normalized Zeeman splitting $\gamma_Z$ are
suggested, and the energy is then determined through
\eqnref{eq:E2}. \Figref{fig:9} displays the variation of transmission
amplitudes with $n=6$ in \eqnref{eq:ringlevel2}. At the point 1 $(f = f_1)$,
the two conditions, \eqnsref{eq:E2} and (\ref{eq:ringlevel2}) are exactly
satisfied with $k_+^+ = \gamma_Z$ so that $T_{+-}$ is almost at its maximum and
the other amplitudes are negligible. Hence, the \textit{conditional spin
  switch} is embodied: the spin $+$ is completely blocked while spin $-$ is
completely flipped. At the same time, the maximal current polarization shown in
\figref{fig:9}(c) indicates that it can also work as the perfect spin polarizer
for unpolarized injection. The opposite spin switch that flips spin $+$ to $-$
can be implemented by reversing the direction of the external magnetic field so
that the two conditions are satisfied with $k_+^- = \gamma_Z < 0$. Note that
the behavior as the perfect spin switch or spin polarizer appears at $f \approx
f_1 \pm 1$ (points 2 and 3) as wells. It is due to the small ratio $g^* m^*/m =
0.1$ used in calculations: $\gamma_Z$ does not change so much for a few periods
of $f$ so that the conditions, \eqnsref{eq:E2} and (\ref{eq:ringlevel2}) are
approximately satisfied at several values of $f$.

The spin flip occurring at the junction discussed in the previous section would
spoil the spin switch efficiency by inducing the tunneling to the other spin
branch, and the spin tunneling cannot be determined only by the spin texture of
the levels in the ring. However, we numerically confirmed that the observed
spin-switch functionality is immune to the variation of $\epsilon$ and $U_b$ as
long as the effective control parameters $\epsilon_\mu$ are small enough. In
fact, the spin flip is very weak if the injection energy is well above the band
bottom $E_+$ [see \figref{fig:3}]. This is the case for \eqnref{eq:E2} as long
as $\gamma_R$ is large enough. One can then safely use the usual analysis of
spin transport based on the spin precession in the ring with no spin flip at
junctions.

\subsubsection{Strong-Coupling Limit}

\begin{figure}[!t]
  \centering
  \includegraphics[width=8.5cm]{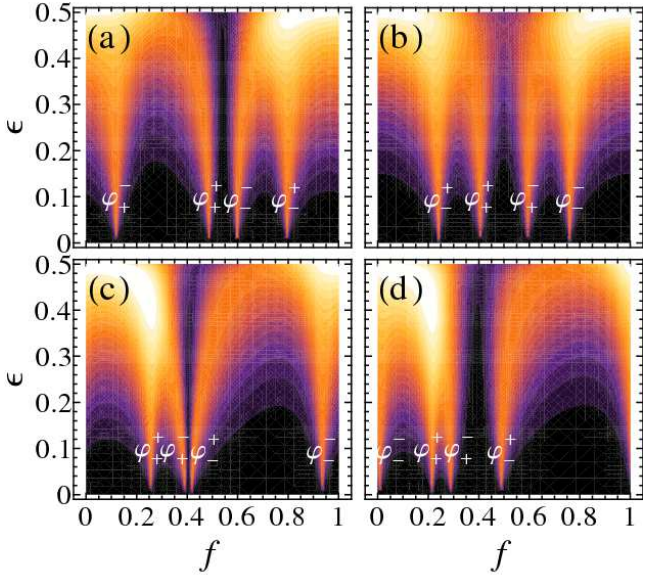}
  \caption{(color online) Contour plots of charge conductance $G$ in unit of
    $e^2/h$ as a function of $f$ and $\epsilon$ for (a) $\gamma_R = \gamma_Z =
    0.4$ and $E = 2 E_+(\gamma_R=0.4,\gamma_Z=0.8)$ (refer to \figref{fig:7})
    and (b,c,d) $\gamma_R = 0.6$ and $E = 2 E_+(\gamma_R=0.6,\gamma_Z=1.3)$
    with $\gamma_Z = 0$ [(b)], 0.7 [(c)], and 1 [(d)]. Here we have used $U_b =
    0$ and the color scale is same as in \figref{fig:7}.}
  \label{fig:10}
\end{figure}

\begin{figure}[!t]
  \centering
  \includegraphics[width=8cm]{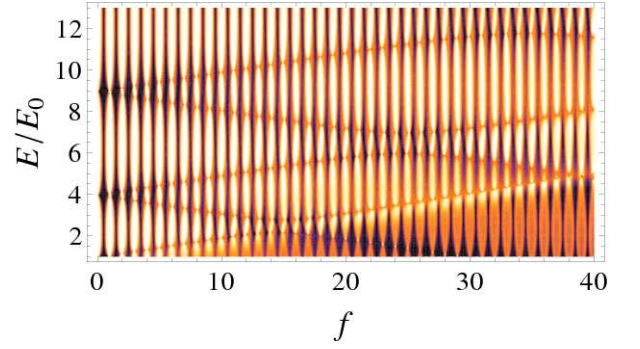}
  \caption{(color online) Contour plot of charge conductance $G$ in unit of
    $e^2/h$ as a function of $f$ and $E/E_0$ in the strong-coupling limit
    $(\epsilon=1/2)$. We have used $U_b = 0$ and $\gamma_R = 0.4$ and the
    Zeeman splitting $\gamma_Z$ increases linearly with $f$: $\gamma_Z =
    0.1\times f$. The color scale is same as in \figref{fig:7}.}
  \label{fig:11}
\end{figure}

\Figref{fig:10} shows the evolution of the charge conductance $G$ as the
lead-ring junction gets more transparent. The resonance feature due to the
ring level, though getting smeared out with increasing $\epsilon$, is still
visible up to $\epsilon \sim 0.4$. For larger values of $\epsilon\gtrsim0.4$,
the conductance peak position does not follow the quantization condition,
\eqnref{eq:ringlevel} any longer, and instead every four consecutive peaks in a
period of $f$ are merged to a single one which is located close to $f =
n\pi$. In addition, a dip is formed between them. The dip appears between the
time-reversal pair states if they are in succession, as can be seen in
\figref{fig:10} (a), (c), and (d). Interestingly, the anti-crossing-like
behavior can be intensified as $\epsilon$ increases as seen in \figref{fig:10}
(c), if the pair states are close to each other in the weak-coupling limit. In
this case the transparent junction enhances the destructive interference
between two resonant levels. The dip can also be formed in other places if the
time-reversal pair is not in succession [see \figref{fig:10} (b)]. In this
case, the dip is less prominent, implying the destructive interference is not
strong enough.

\begin{figure}[!t]
  \centering
  \includegraphics[width=8cm]{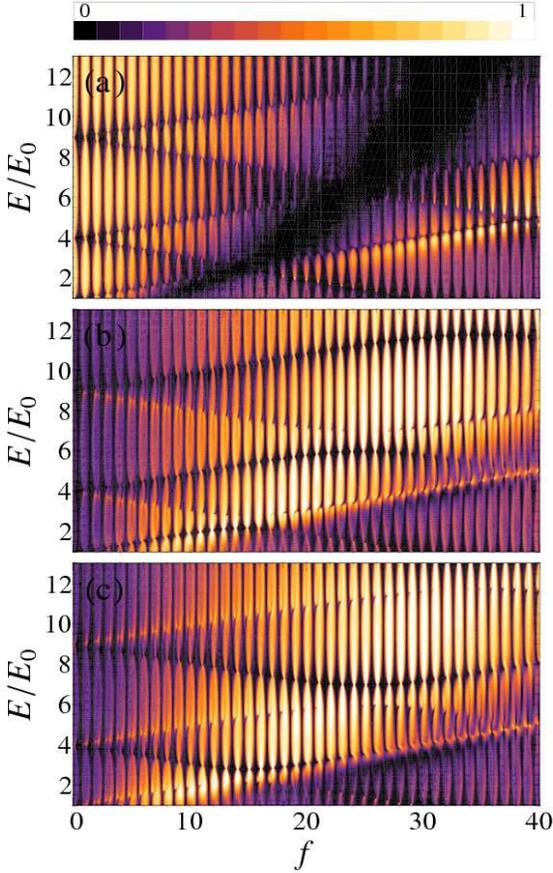}
  \caption{(color online) Contour plot of (a) spin-conserving transmission
    amplitudes $T_{++} = T_{--}$ and (b,c) spin-flip transmission amplitudes
    $T_{-+}$ [(b)] and $T_{+-}$ [(c)] as functions of $f$ and $E/E_0$ in the
    strong-coupling limit $(\epsilon=1/2)$. Here the polarization axis of two
    leads are chosen to align with the positive $x$ axis:
    $(\vartheta_\ell,\varphi_\ell) = (\pi/2,0)$. Values of other parameters are
    same as in \figref{fig:11}.}
  \label{fig:12}
\end{figure}

The charge transport in the strong-coupling limit $(\epsilon \sim 1/2)$, as
seen in \figref{fig:11}, clearly exhibits the well-known AB oscillations as the
magnetic flux is varied. In addition, the Zeeman splitting $\gamma_Z$,
increasing linearly with $f$, superposes line-shaped patterns upon the AB
oscillations along which the conductance is suppressed. This suppression is due
to the localization effect in the ring. To be simple, consider the Rashba-free
system. The analytical expression for spin-dependent transmission amplitude is
then available:
\begin{align}
  T_\mu
  =
  \frac{4\epsilon_\mu^{\prime2} \cos^2\pi f \sin^2\pi \widetilde{k}_\mu}
  {\left|
      \epsilon'_\mu e^{2\pi i\widetilde{k}_\mu} - \cos 2\pi\widetilde{k}_\mu
      + \left(\frac{1-p_\mu}{2}\right)^2
      + \left(\frac{1+p_\mu}{2}\right)^2 \cos2\pi f
    \right|^2}
\end{align}
with $\epsilon'_\mu$ given by \eqnref{eq:e}, $p_\mu = \sqrt{1 -
  2\epsilon'_\mu}$ and $\widetilde{k}_\mu = \sqrt{E/E_0 - \mu\gamma_Z}$. The
transmission vanishes not only when $f = n + 1/2$ but also when
$\widetilde{k}_\mu = n$ where $n$ is an integer. The latter condition means
that the wave in the ring forms the standing wave so that the state is
localized and does not contribute to the transport. Hence the conductance
suppression happens at $E/E_0 = n^2 + \mu\gamma_Z$, making spin-dependent dark
lines in the charge conductance [see \figref{fig:11}]. The Rashba SOC, present
in our system but rather small, makes a perturbative coupling between
spin-$\up$ and $\down$ states, inducing the anti-crossing of dark lines that
would be degenerate otherwise.
Finally, one can notice that in the lower right corner of \figref{fig:11}
(under the line $E/E_0 = 1 + \gamma_Z$) the charge conductance is quite
suppressed; the maximum is reduced by half, reaching $e^2/h$, not $2e^2/h$. It
is because in this region $E < E_+$ so that only the spin-$-$ channel is
open. The spin-$+$ channel exists in the evanescent waves whose contribution
decreases exponentially with $E_+-E$.

\begin{figure}[!t]
  \centering
  \includegraphics[width=8cm]{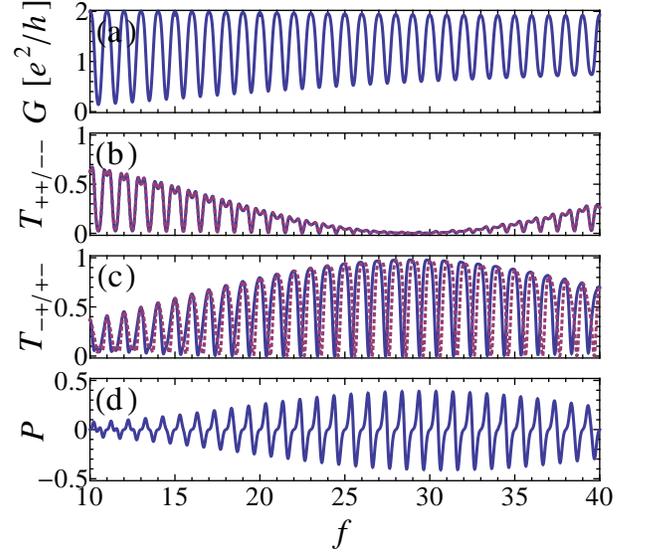}
  \caption{(color online) (a) Charge conductance $G$, (b) spin-conserving
    transmission amplitudes $T_{++}$ (solid line), $T_{--}$ (dotted line), (c)
    spin-flip transmission amplitudes $T_{-+}$ (solid line), $T_{+-}$ (dotted
    line), and (d) current polarization as functions of $f$ along the $E =
    9E_0$ line in \figref{fig:12}. Here the polarization axis of two leads are
    chosen to align with the positive $x$ axis: $(\vartheta_\ell,\varphi_\ell)
    = (\pi/2,0)$. Values of other parameters are same as in \figref{fig:12}.}
  \label{fig:13}
\end{figure}

The spin transport in the strong-coupling limit is examined in \figref{fig:12}.
Similarly to the charge conductance, the spin-dependent transmissions feature
the AB oscillations and the localization-induced dark line patterns. In
addition, they also exhibit a global modulation of the height of the AB
peaks. Interestingly, the modulation patterns are in opposite trends between
spin-conserving transmissions ($T_{++}$ and $T_{--}$) and spin-flip
transmissions ($T_{+-}$ and $T_{-+}$): when the spin-conserving transmissions
are strong the spin-flip transmissions are weak and vice versa. This opposing
behavior is clearly displayed in \figref{fig:13} where the charge and spin
transmissions are calculated at a given injection energy, $E = 9E_0$. This
global modulation of spin-dependent transmission is surely related to the
variation of $\gamma_Z$ with $f$: $\gamma_Z = 0.1\times f$ is used
here. Subsequently, the non-adiabatic geometric phase connected to the Rashba
SOC and the Zeeman splitting varies gradually and changes the interference
between the ring modes, resulting in the modulation of the spin-dependent
transmission. With the total charge transmission unchanged so much, the
decrease of the spin-conserving transmission then accompanies the enhancement of
the spin-flip transmission. Hence, in the regime of parameters where the
spin-conserving transmissions are negligible, a \textit{unconditional spin
  switch} is implemented: the injected spin $+$ is switched to the spin $-$ and
vice versa. As can be seen in \figref{fig:12} and \figref{fig:13}, the
parameter regime for the system to act as a good spin switch is quite wide: the
working condition encloses several periods of $f$ and wide range of energy. It
is attributed to the slow variation of the geometrical phase with $f$. Finally,
this system can also behavior as a good spin polarizer for unpolarized current
injection, as seen in \figref{fig:13}(d). Since the maxima of $T_{-+}$ and
$T_{+-}$ are off the synchronization, the current polarization oscillates
strongly between -0.4 and 0.4. The polarization of spin current can then be
easily tuned by changing the magnetic flux by the half flux quantum $\Phi_0/2$.

\section{Discussion and Conclusion\label{sec:discussion}}

We have proposed a general scattering-matrix formalism that naturally
guarantees the charge conservation through a quantum ring with arbitrary
spin-dependent interactions. To the end, we insert artificial SOC-free buffers
in the vicinity of every junction and solve the system Hamiltonian in a
standard way. The original problem is recovered by shrinking the size of
buffers to zero, while the effect of buffers still remains. It is found that as
long as the ring has nonorthogonal spin textures the spin-flip scattering can
happen even if the junction itself is nonmagnetic. In the case of $n$-type
semiconductor with both the Rashba SOC and the Zeeman splitting, the finite
spin-flip scattering and the conservation of charge current are numerically
confirmed. In addition, it is found that the interplay of the AB and AC
effects, in the presence of the Zeeman splitting, enables the ring
interferometer to act as conditional/unconditional spin switch in the
weak/strong coupling limit.

It should be noted that our formalism is not restricted to the structure of the
AB interferometer used in this paper. The technique of inserting artificial
buffers and shrinking them to zero can be applied to any network of
semiconductors with arbitrary SOC. As stated above, the merit of our formalism
is that the charge conservation at junctions is guaranteed as long as the
interfaces between buffers and the spin-dependent regions are treated
correctly.

While in our study we focus on the simplest scattering matrix by minimizing the
number of physical parameters for buffers, the scattering matrix can be more
generalized by introducing some spin-dependent coupling into the buffers in a
controlled way. The extended form of the scattering matrix may give us a hint
for the general structure of the scattering matrix connecting any
spin-dependent channels with a single constraint: the charge current
conservation. It would be interesting to find out the general form of the
scattering-matrix based on no other than the conservation law without leaning
on the specific model such as buffers.

\acknowledgements

This work is supported by grants from the Kyung Hee University Research Fund
(KHU-20090742).

\appendix


\begin{thebibliography}{99}
\bibitem{eigenstate_validity} The solution, \eqnref{eq:evec} is valid only when
  the energy $E$ is larger or equal to the bottom of the upper spin branch,
  $E_+$. Otherwise, the evanescent waves should be taken into account or all
  the four eigenstates belong to the lower spin branch.
\end{thebibliography}
\end{document}